\documentclass[12pt,a4paper]{article}
\usepackage{a4wide}
\usepackage{amsmath}
\usepackage{amssymb}
\usepackage{epsfig}
\usepackage{subfigure}
\usepackage{exscale}
\usepackage{float}
\usepackage{bbm}
\usepackage[numbers,sort&compress]{natbib}
\usepackage{pst-plot, pstricks,pst-math}
\usepackage{fancybox,amssymb,color}
\usepackage{graphicx}
\usepackage{pstricks, color, graphicx, epsfig, psfrag}
\usepackage{amsfonts,amsmath,amssymb,slashed}
\usepackage{dsfont}
\usepackage{bbm,bm}
\usepackage{fancyhdr, a4wide}
\usepackage[english]{babel}
\usepackage{subfigure}
\makeatletter
\@addtoreset{equation}{section}
\makeatother

\title{Chiral Perturbation Theory Analysis of the Quark Condensate in a Magnetic Field}

\author{Christoph P.\ Hofmann$^a$ \\ \\
\normalsize{$^a$ Facultad de Ciencias, Universidad de Colima} \\
\vspace{0.3cm}
\normalsize {Bernal D\'iaz del Castillo 340, Colima C.P.\ 28045, Mexico} \\}

\begin{document}

\maketitle

\begin{abstract} \normalsize

We present two-loop results for the quark condensate in an external magnetic field within chiral perturbation theory using coordinate space
techniques. At finite temperature, we explore the impact of the magnetic field on the pion-pion interaction in the quark condensate for
arbitrary pion masses and derive the correct weak magnetic field expansion in the chiral limit. At zero temperature, we provide the
complete two-loop representation for the vacuum energy density and the quark condensate.

\end{abstract}

\maketitle

\section{Introduction}
\label{Intro}

The quark condensate -- order parameter of spontaneous chiral symmetry breaking -- is a crucial quantity in particle physics. It comes with
no surprise that the relevant literature is extensive. Here we focus on the properties of the quark condensate in an external constant
magnetic field. Our calculation within the framework of chiral perturbation theory (CHPT) goes up to two-loop order, but in contrast to the
available CHPT-studies -- see Refs.~\citep{SS97,AS00,Aga00,Aga01,AS01,CMW07,Aga08,Wer08,And12a,And12b} -- we use a coordinate-space
representation for the pion propagators and the associated kinematical functions. Other references, also dealing with the quark condensate
in a magnetic field, are based upon lattice QCD \citep{EMS10,EN11,BBEFKKSS12a,BBKKP12,BBEFKKSS12b,BBCCEKPS12c,BBCKS14,IMPS14,EMNS18,
EGKKP19}, feature analytical studies relying on the Nambu-Jona-Lasinio model \citep{GR11,AA13,FCMPS14,FCLFP14,FCP14,FIPQ14,ZFL16}, or
comprise yet other models and methods \citep{NK11,FR11,BEK13,OS13,CFS14,OS14a,HPS14,MP15}.

In a recent article, Ref.~\citep{Hof19}, the present author has pointed out that -- in the chiral limit -- the two published one-loop
series for the finite-temperature quark condensate in a weak magnetic field, independently derived by different authors, are erroneous. The
proper series at one-loop order has been established in Ref.~\citep{Hof19} -- one of our goals in the actual study is to review the
situation at two-loop order. Indeed, errors also occur here. We clarify the situation by providing the correct weak magnetic field
expansion of the finite-temperature quark condensate in the chiral limit. One of the advantages of our coordinate-space approach is that it
allows for a transparent derivation of the various limits that have to be taken in the calculation: chiral limit ($M \to 0$) and weak
magnetic field limit ($|qH| \ll T^2$).

Apart from straightening these issues, we also investigate the impact of the magnetic field on the pion-pion interaction in the quark
condensate for arbitrary pion masses. At finite temperature, the interaction constitutes up to ten percent as compared to the leading
noninteracting pion gas contribution, and is most pronounced in the chiral limit. When the magnetic field increases, the finite-temperature
quark condensate (sum of one- and two-loop contribution at fixed temperature and pion mass) grows monotonically. The effect is again most
pronounced in the chiral limit.

Using the dressed pions as pertinent degrees of freedom, the low-temperature series of the quark condensate is characterized by a
$T^2$-contribution that refers to the dressed but non-interacting pions, while interaction effects emerge at order $T^4$. In the chiral
limit and in weak magnetic fields, the series at {\bf order $\mathbf T^2$} -- organized by the expansion parameter $\epsilon = |qH|/T^2$
($q$ is the electric charge of the pion) -- involves a leading square-root term $\propto \sqrt{\epsilon}$, a term linear in $\epsilon$,
followed by a half-integer power $\epsilon^{3/2}$ and a logarithmic contribution $\epsilon^2 \ln \epsilon$. The remaining contributions
involve even powers of $\epsilon$. At {\bf order $\mathbf T^4$} the series exhibits the same structure, with the exception that a term
linear in $\epsilon$ is absent -- in contrast to what has been reported in the literature.

Finally, we provide the two-loop representation for the QCD vacuum energy density and the zero-temperature quark condensate. The
representation involves nonanalytic contributions in the form of logarithms, as well as Gamma and Polygamma functions that depend
nontrivially on the ratio between magnetic field and pion mass. In contrast to previous studies we provide the full two-loop representation
-- not merely the terms that are induced by the nonzero magnetic field.

The article is organized as follows. The two-loop CHPT evaluation is briefly reviewed in Sec.~\ref{CHPT} to set the basis for the
subsequent analysis. In Sec.~\ref{quarkCondensate} we explore the quark condensate at finite and zero temperature for arbitrary pion masses
-- in particular also for the physical pion masses -- in presence of a magnetic field. In the same section we furthermore compare our
findings with the literature and point out errors in the published results. Finally, Sec.~\ref{conclusions} contains our conclusions. More
technical issues are presented in three appendices. In Appendix \ref{appendixA} we discuss in detail the two-loop CHPT evaluation at zero
temperature. While Appendix \ref{appendixB} is devoted to the chiral limit in nonzero magnetic fields at $T$=0, in Appendix \ref{appendixC}
we consider the same situation at finite temperature which boils down to an analysis of the various kinematical functions required.

\section{Chiral Perturbation Theory Evaluation}
\label{CHPT}

\begin{figure}
\begin{center}
\includegraphics[width=15cm]{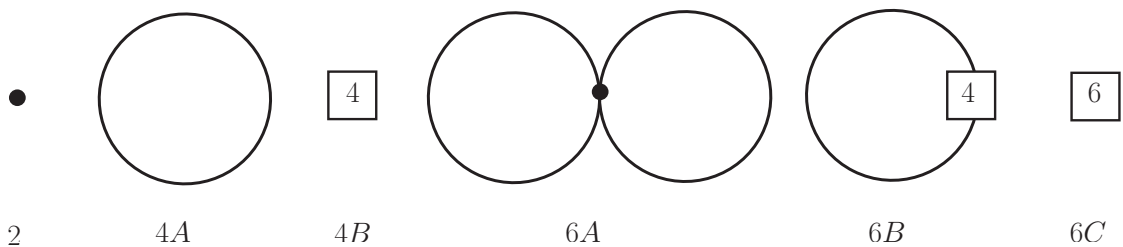}
\end{center}
\caption{Chiral perturbation theory diagrams for the QCD free energy density up to order $p^6$. Vertices from ${\cal L}^2_{eff}$ (filled
circles), as well as vertices from ${\cal L}^4_{eff}$ and ${\cal L}^6_{eff}$ (denoted by the numbers $4$ and $6$) contribute. The lines refer
to the thermal pion propagators.}
\label{figure1}
\end{figure}

The relevant low-energy excitations in two-flavor chiral perturbation theory\footnote{For reviews of chiral perturbation theory see, e.g.,
Refs.~\citep{Leu95,Sch03}.} are the three pions that are incorporated in the SU(2) matrix $U(x)$ as
\begin{equation}
U(x) =\exp(i \tau^i \pi^i(x)/F) \, , \qquad i=1,2,3 \, .
\end{equation}
Here $\tau^i$ are the Pauli matrices and $F$ stands for the tree-level pion decay constant. While $\pi^0$ describes the neutral
pion\footnote{Although the Pauli matrix associated with the neutral pion is $\tau^3$, we will denote the neutral pion field as $\pi^0$ in
view of its zero charge.}, the charged pions correspond to the linear combinations
\begin{equation}
\pi^\pm = \frac{1}{\sqrt{2}} \Big( \pi^1 \pm i\pi^2 \Big) \, . 
\end{equation}
The Euclidean leading-order (order $p^2$) effective Lagrangian is given by
\begin{equation}
\label{L2}
{\cal L}^2_{eff} = \mbox{$ \frac{1}{4}$} F^2 Tr \Big[ {(D_{\mu} U)}^\dagger (D_{\mu} U) - M^2 (U + U^\dagger) \Big] \, ,
\end{equation}
where $M$ is the tree-level pion mass.  In the covariant derivative,
\begin{equation}
D_{\mu} U = \partial_\mu U + i [Q,U] A^{EM}_\mu \, ,
\end{equation}
the quantity $Q$ is the charge matrix of the quarks, i.e., $Q=diag(2/3,-1/3)e$, while the magnetic field $H$ enters via the gauge field
$A^{EM}_\mu=(0,0,-H x_1,0)$. As illustrated in Fig.~\ref{figure1}, a two-loop calculation of the free energy density in addition involves
the subleading pieces ${\cal L}^4_{eff}$ and ${\cal L}^6_{eff}$ of the effective Lagrangian.

The set of terms proportional to four pion fields generated by the leading piece ${\cal L}^2_{eff}$ -- as required for the evaluation of the
two-loop diagram 6A -- are
\begin{eqnarray}
{\cal L}^2_{\{4\}} & = & \frac{1}{3 F^2} \, \pi^0 \partial_{\mu} \pi^0 \Big( \partial_{\mu} \pi^+ \pi^- + \partial_{\mu} \pi^- \pi^+ \Big)
- \frac{1}{3 F^2} \, \partial_{\mu} \pi^0 \partial_{\mu} \pi^0 \pi^+ \pi^- \nonumber \\
& & - \frac{1}{3 F^2} \, \pi^0 \pi^0 \partial_{\mu} \pi^+ \partial_{\mu} \pi^-
- \frac{1}{3 F^2} \, \pi^+ \pi^- \partial_{\mu} \pi^+ \partial_{\mu} \pi^-  \nonumber \\
& & + \frac{1}{6 F^2} \, \Big(\partial_{\mu} \pi^+ \pi^- \partial_{\mu} \pi^+ \pi^- + \partial_{\mu} \pi^- \pi^+ \partial_{\mu} \pi^- \pi^+\Big)
\, .
\end{eqnarray}
Other pieces from ${\cal L}^2_{eff}$ needed for our calculation are terms with two (diagram 4A) or zero (diagram 2) pion fields,
\begin{eqnarray}
{\cal L}^2_{\{2\}} & = & \mbox{$ \frac{1}{2}$} \partial_{\mu} \pi^0 \partial_{\mu} \pi^0 + \partial_{\mu} \pi^+ \partial_{\mu} \pi^-
+ \mbox{$ \frac{1}{2}$} M^2 \pi^0 \pi^0 + M^2 \pi^+ \pi^- \, , \nonumber \\
{\cal L}^2_{\{0\}} & = & - F^2 M^2 \, .
\end{eqnarray}

As for the subleading piece ${\cal L}^4_{eff}$, we use the representation given in Eq.~(D.2) of Ref.~\citep{Sch03}. The relevant terms for
our calculation are those that contain two (diagram 6B) or zero (diagram 4B) pion fields, 
\begin{eqnarray}
{\cal L}^4_{\{2\}} & = & l_3 \frac{M^4}{F^2} \, \pi^0 \pi^0 + 2l_3 \frac{M^4}{F^2} \, \pi^+ \pi^-
+ (4l_5 - 2l_6 ) \frac{{|qH|}^2}{F^2} \, \pi^+ \pi^- \, , \nonumber \\
{\cal L}^4_{\{0\}} & = & -(l_3 + h_1) M^4 + 4 h_2 {|qH|}^2 \, .
\end{eqnarray}
The quantities $l_3, l_5, l_6, h_1, h_2$ are next-to-leading order (NLO) low-energy effective constants.

Finally, following Ref.~\citep{BCE00}, the terms from ${\cal L}^6_{eff}$ contributing to the tree-level diagram 6C read
\begin{equation}
{\cal L}^6_{\{0\}} = - 16 (c_{10} + 2 c_{11}) M^6 - 8 c_{34} M^2 |qH|^2 \, ,
\end{equation}
where $c_{10}, c_{11}, c_{34}$ are  next-to-next-to-leading order (NNLO) low-energy effective constants.

It is convenient to divide the free energy density into two pieces as
\begin{equation}
z = z_0 + z^T \, ,
\end{equation}
where $z_0$ contains all $T$=0 contributions (vacuum energy density), and $z^T$ involves the finite-temperature part -- both terms depend
on the magnetic field. Before addressing the $T$=0 case, we quote the result for the finite-temperature piece which has been derived within
the CHPT coordinate-space approach up to two-loop order in Ref.~\citep{Hof20}:
\begin{eqnarray}
\label{fedPhysicalM}
z^T & = & - g_0(M^{\pm}_{\pi},T,0) -\mbox{$ \frac{1}{2}$} g_0(M^0_{\pi},T,0)- {\tilde g}_0(M^{\pm}_{\pi},T,H) \nonumber \\
& & + \frac{M^2_{\pi}}{2 F^2} \, g_1(M^{\pm}_{\pi},T,0) \, g_1(M^0_{\pi},T,0)
- \frac{M^2_{\pi}}{8 F^2} \, {\Big\{ g_1(M^0_{\pi},T,0)  \Big\}}^2 \nonumber \\
& & + \frac{M^2_{\pi}}{2 F^2} \, g_1(M^0_{\pi},T,0) \, {\tilde g}_1(M^{\pm}_{\pi},T,H) + {\cal O}(p^8) \, .
\end{eqnarray}
The kinematical Bose functions are defined as 
\begin{eqnarray}
\label{boseFunctions}
g_0({\cal M},T,0) & = & T^4 \, {\int}_{\!\!\! 0}^{\infty}  \mbox{d} \rho \rho^{-3} \, \exp\Big( -\frac{{\cal M}^2}{4 \pi T^2}
\rho \Big) \Bigg[ S\Big( \frac{1}{\rho} \Big) -1 \Bigg] \, , \nonumber \\
g_1({\cal M},T,0) & = & \frac{T^2}{{4 \pi}} \, {\int}_{\!\!\! 0}^{\infty}  \mbox{d} \rho \rho^{-2} \, \exp\Big( -\frac{{\cal M}^2}{4 \pi T^2}
\rho \Big) \Bigg[ S\Big( \frac{1}{\rho} \Big) -1 \Bigg] \, , \nonumber \\
{\tilde g}_0(M^{\pm}_{\pi},T,H) & = & \frac{T^2}{{4 \pi}} \, |qH| {\int}_{\!\!\! 0}^{\infty} \mbox{d} \rho \rho^{-2} \,
\Bigg( \frac{1}{\sinh(|qH| \rho /4 \pi T^2)} - \frac{4 \pi T^2}{|qH| \rho} \Bigg) \nonumber \\
& & \times \, \exp\Big( -\frac{{(M^{\pm}_{\pi})}^2}{4 \pi T^2} \rho \Big) \Bigg[ S\Big( \frac{1}{\rho} \Big) -1 \Bigg] \nonumber \\
{\tilde g}_1(M^{\pm}_{\pi},T,H) & = & \frac{1}{16 \pi^2} \, |qH| {\int}_{\!\!\! 0}^{\infty} \mbox{d} \rho \rho^{-1} \,
\Bigg( \frac{1}{\sinh(|qH| \rho /4 \pi T^2)} - \frac{4 \pi T^2}{|qH| \rho} \Bigg) \nonumber \\
& & \times \, \exp\Big( -\frac{{(M^{\pm}_{\pi})}^2}{4 \pi T^2} \rho \Big) \Bigg[ S\Big( \frac{1}{\rho} \Big) -1 \Bigg] \, ,
\end{eqnarray}
and $S(z)$ stands for the Jacobi theta function,
\begin{equation}
S(z) = \sum_{n=-\infty}^{\infty} \exp(- \pi n^2 z) \, .
\end{equation}
Note that ${\tilde g}_0$ and ${\tilde g}_1$ explicitly depend on the magnetic field through the hyperbolic sine and that they involve the
mass $M^{\pm}_{\pi}$, i.e., the masses of the charged pions in a magnetic field given by
\begin{equation}
\label{chargedPionMass}
{(M^{\pm}_{\pi})}^2 = M^2_{\pi} + \frac{{\overline l}_6 - {\overline l}_5}{48 \pi^2} \, \frac{{|qH|}^2}{F^2} \, .
\end{equation}
The mass $\cal M$ in $g_0$ and $g_1$, according to Eq.~(\ref{fedPhysicalM}), can either represent $M^{\pm}_{\pi}$ or $M^0_{\pi}$, where the
latter is the  mass of the neutral pion in a magnetic field,
\begin{equation}
\label{neutralPionMass}
{(M^0_{\pi})}^2 = M^2_{\pi}  + \frac{M^2}{F^2} \, K_1 \, ,
\end{equation}
and $K_1$ denotes the integral
\begin{equation}
\label{intK1}
K_1 = \frac{|qH|}{16 \pi^2}  \, {\int}_{\!\!\! 0}^{\infty} \mbox{d} \rho \, \rho^{-1} \, \exp\Big( -\frac{M^2_{\pi}}{|qH|} \rho \Big) \,
\Big( \frac{1}{\sinh(\rho)} - \frac{1}{\rho} \Big) \, .
\end{equation}
The kinematical functions $g_0$ and $g_1$ hence implicitly depend on the magnetic field through the neutral and charged pion masses.
Finally, the mass $M_{\pi}$ is the renormalized NLO pion mass in zero magnetic field,
\begin{equation}
\label{Mpi}
M^2_{\pi} = M^2 - \frac{{\overline l}_3}{32 \pi^2} \, \frac{M^4}{F^2} + {\cal O}(M^6) \, .
\end{equation}
The quantities ${\overline l}_3, {\overline l}_5, {\overline l}_6$ are renormalized NLO low-energy effective constants -- details are
provided in Appendix \ref{appendixA1}.

We now address the zero-temperature part in the free energy density\footnote{To the best of our knowledge, the complete CHPT two-loop
representation for the QCD vacuum energy density -- containing magnetic-field dependent as well as $H$-independent terms -- is not
available in the literature.}. Apart from the temperature-independent tree-level graphs 2, 4B and 6C, we also have $T$=0 contributions from
the loop graphs. This is because the thermal propagators for the pions,
\begin{eqnarray}
\label{ThermalPropagator}
G^{\pm}(x) & = & \sum_{n = - \infty}^{\infty} \Delta^{\pm}({\vec x}, x_4 + n \beta) \, , \nonumber \\
G^0(x) & = & \sum_{n = - \infty}^{\infty} \Delta^0({\vec x}, x_4 + n \beta) \, , \qquad \beta = \frac{1}{T} \, ,
\end{eqnarray}
contain a zero-temperature piece associated with $n$=0. In Appendix \ref{appendixA2} we process these $T$=0 contributions and show that all
UV-divergences cancel. The final result for the renormalized the vacuum energy density at order $p^6$ then amounts to
\begin{eqnarray}
\label{freeEDp6ZeroT}
z^{[6]}_0 & = & \frac{3{\overline l}_3 ({\overline c}_{10} + 2 {\overline c}_{11})}{1024 \pi^4} \, \frac{M^6}{F^2}
- \frac{({\overline l}_6 - {\overline l}_5){\overline c}_{34}}{768 \pi^4} \, \frac{{|qH|}^2 M^2}{F^2} \nonumber \\
& & - \frac{{\overline l}_3}{32 \pi^2} \,  \frac{M^4}{F^2} \, K_1
+ \frac{({\overline l}_6 - {\overline l}_5)}{48 \pi^2} \, \frac{{|qH|}^2}{F^2} \, K_1 \, .
\end{eqnarray}
The quantities ${\overline l}_i$ and ${\overline c}_i$ are the renormalized NLO and NNLO effective constants, defined in Appendix
\ref{appendixA1}.

The full vacuum energy density also includes the zero-temperature pieces of order $p^4$ and $p^2$,
\begin{equation}
z_0 = z^{[6]}_0 + z^{[4]}_0 + z^{[2]}_0 \, ,
\end{equation}
which are (see Ref.~\citep{Hof19} for $z^{[4]}_0$), 
\begin{eqnarray}
\label{freeEDp4ZeroT}
z^{[4]}_0 & = & \frac{M^4}{64 \pi^2} \, \Big({\overline l_3} - 4{\overline h_1} - \frac{3}{2}\Big) + \frac{{|qH|}^2}{96 \pi^2} \,
( {\overline h_2} - 1)  \nonumber \\
& & - \frac{{|qH|}^2}{16 \pi^2} {\int}_{\!\!\! 0}^{\infty} \mbox{d} \rho \rho^{-2} \Big( \frac{1}{\sinh(\rho)} - \frac{1}{\rho} + \frac{\rho}{6}
\Big) \, \exp\!\Big( -\frac{M^2}{|qH|} \rho \Big) \, , \nonumber \\
z^{[2]}_0 & = & - F^2 M^2 \, .
\end{eqnarray}
The subleading contributions $z^{[4]}_0$ and $z^{[6]}_0$ as displayed above, i.e., the renormalized expressions, are independent of the
renormalization scale $\mu$. This is a nontrivial consistency check of our calculation. We now turn to the quark condensate which is the
main subject of the present investigation.

\section{Quark Condensate in a Magnetic Field}
\label{quarkCondensate}

The quark condensate is given by the derivative of the free energy density with respect to the quark mass\footnote{Throughout the study we
work in the isospin limit $m = m_u = m_d$.}
\begin{equation}
\langle {\bar q} q \rangle = \frac{\mbox{d} z}{\mbox{d} m} \, .
\end{equation}
At zero temperature it corresponds to the vacuum expectation value
\begin{equation}
\langle 0 | {\bar q} q | 0 \rangle = \frac{\mbox{d} z_0}{\mbox{d} m}
= -\frac{{\langle 0 | {\bar q} q | 0 \rangle}_0}{F^2} \, \frac{\mbox{d} z_0}{\mbox{d} M^2} \, .
\end{equation}
Note that we have used the leading-order Gell-Mann--Oakes--Renner relation
\begin{equation}
M^2 = -\frac{m}{F^2 } \, {\langle 0 | {\bar q} q | 0 \rangle}_0 \, ,
\end{equation}
where the quantity ${\langle 0 | {\bar q} q | 0 \rangle}_0$ is the quark condensate at  $T$=0 (and zero magnetic field) in the chiral
limit -- as indicated by the lower index "0". The purely finite-temperature part in the quark condensate amounts to
\begin{equation}
{\langle {\bar q} q \rangle}^T = - \frac{\mbox{d} P}{\mbox{d} m}
= \frac{{\langle 0 | {\bar q} q | 0 \rangle}_0}{F^2} \, \frac{\mbox{d} P}{\mbox{d} M^2} \, .
\end{equation}
Up to the sign, the pressure is nothing but the finite-temperature piece in the free energy density,
\begin{equation}
P = -z^T \, .
\end{equation}
In the representation of $z^T$, Eq.~(\ref{fedPhysicalM}), we have used the NLO renormalized pion mass $M_{\pi}$ instead of $M$. The
connection between the two quantities is given by Eq.~(\ref{Mpi}). For the quark condensate we then obtain
\begin{equation}
\langle {\bar q} q \rangle = \frac{{\langle 0 | {\bar q} q | 0 \rangle}_0}{F^2} \,
\Bigg\{\!  -\frac{\mbox{d} z_0}{\mbox{d} M^2_{\pi}} + \frac{\mbox{d} P}{\mbox{d} M^2_{\pi}} \Bigg\} \Bigg( 1 - \frac{ M^2_{\pi}}{32 \pi^2 F^2} \,
(2 {\overline l}_3 -1) \Bigg) \, .
\end{equation}
In the parenthesis we have replaced $M^2$ by $M^2_{\pi}$ which is legitimate at the order we are operating.

It should be pointed out that the zero-temperature quark condensate at order $p^4$, according to Eq.~(\ref{freeEDp4ZeroT}), involves the
NLO effective constant $\overline h_1$ which depends on the renormalization convention (see Ref.~\citep{GL84}). No such ambiguities due to
NLO effective constants $\overline h_i$ are introduced in the zero-temperature quark condensate at order $p^6$, according to
Eq.~(\ref{freeEDp6ZeroT}). Likewise, the  finite-temperature part of the quark condensate is also free of such renormalization ambiguities.

\subsection{Finite-Temperature Quark Condensate}

In order to make powers of temperature in the quark condensate manifest, instead of operating with the Bose functions $g_r$ and
${\tilde g}_r$, we now work with the dimensionless functions $h_r$ and ${\tilde h}_r$ defined as
\begin{equation}
\label{conversion}
h_0 = \frac{g_0}{T^4} \, , \quad  {\tilde h}_0 = \frac{{\tilde g}_0}{T^4} \, , \qquad
h_1 = \frac{g_1}{T^2} \, , \quad  {\tilde h}_1 = \frac{{\tilde g}_1}{T^2} \, , \qquad
h_2 = g_2 \, , \quad  {\tilde h}_2 = {\tilde g}_2 \, .
\end{equation}
With the expression for $z^T$, Eq.~(\ref{fedPhysicalM}), the finite-temperature part of the quark condensate takes the form
\begin{equation}
\frac{{\langle {\bar q} q \rangle}^T}{{\langle 0 | {\bar q} q | 0 \rangle}_0} \, {\Bigg( 1 - \frac{ M^2_{\pi}}{32 \pi^2 F^2} \,
(2 {\overline l}_3 -1)\Bigg)}^{-1}
= - \Big\{ \frac{q_1}{F^2} T^2 + \frac{q_2}{F^4} T^4  + {\cal O}(T^6) \Big\} \, .
\end{equation}
The respective coefficients,
\begin{eqnarray}
q_1 & = &  h_1(M^{\pm}_{\pi},T,0) + \mbox{$ \frac{1}{2}$} a_0 h_1(M^0_{\pi},T,0) + {\tilde h}_1(M^{\pm}_{\pi},T,H) \, , \nonumber \\
q_2 & = & + \mbox{$ \frac{1}{2}$} h_1(M^{\pm}_{\pi},T,0) h_1(M^0_{\pi},T,0)
+ \mbox{$ \frac{1}{2}$} h_1(M^0_{\pi},T,0) {\tilde h}_1(M^{\pm}_{\pi},T,H) \nonumber \\
& & - \mbox{$ \frac{1}{8}$} h_1(M^0_{\pi},T,0) h_1(M^0_{\pi},T,0)
- \mbox{$ \frac{1}{2}$} \frac{m^2}{t^2} h_1(M^0_{\pi},T,0) h_2(M^{\pm}_{\pi},T,0) \nonumber \\
& & - \mbox{$ \frac{1}{2}$} a_0 \frac{m^2}{t^2} h_1(M^{\pm}_{\pi},T,0) h_2(M^0_{\pi},T,0)
- \mbox{$ \frac{1}{2}$} a_0 \frac{m^2}{t^2} {\tilde h}_1(M^{\pm}_{\pi},T,H) h_2(M^0_{\pi},T,0) \nonumber \\
& & + \mbox{$ \frac{1}{4}$}a_0 \frac{m^2}{t^2} h_1(M^0_{\pi},T,0) h_2(M^0_{\pi},T,0)
- \mbox{$ \frac{1}{2}$} \frac{m^2}{t^2} h_1(M^0_{\pi},T,0) {\tilde h}_2(M^{\pm}_{\pi},T,H) \, ,
\end{eqnarray}
depend in a nontrivial way on the ratios between pion masses, magnetic field and temperature. The NLO mass correction $a_0$ is
\begin{equation}
a_0 = \frac{\mbox{d} {(M^0_{\pi})}^2 }{\mbox{d} M^2_{\pi}} =  1 + \frac{K_1}{F^2}
+ \frac{M^2_{\pi}}{F^2} \, \frac{\mbox{d} K_1}{\mbox{d} M^2_{\pi}} \, ,
\end{equation}
with the integral $\mbox{d} K_1/\mbox{d} M^2_{\pi}$ given by
\begin{equation}
\frac{\mbox{d} K_1}{\mbox{d} M^2_{\pi}} = - \frac{1}{16 \pi^2} \, {\int}_{\!\!\! 0}^{\infty} \mbox{d} \rho \, \exp\Big( -\frac{M^2_{\pi}}{|qH|}
\rho \Big) \, \Big( \frac{1}{\sinh(\rho)} - \frac{1}{\rho} \Big) \, .
\end{equation}
The coefficient $q_1$ refers to the free pion gas contribution of order $T^2$, while the coefficient $q_2$ captures the pion-pion
interaction that emerges at order $T^4$ in the finite-temperature quark condensate.

\begin{figure}
\begin{center}
\hbox{
\includegraphics[width=8.0cm]{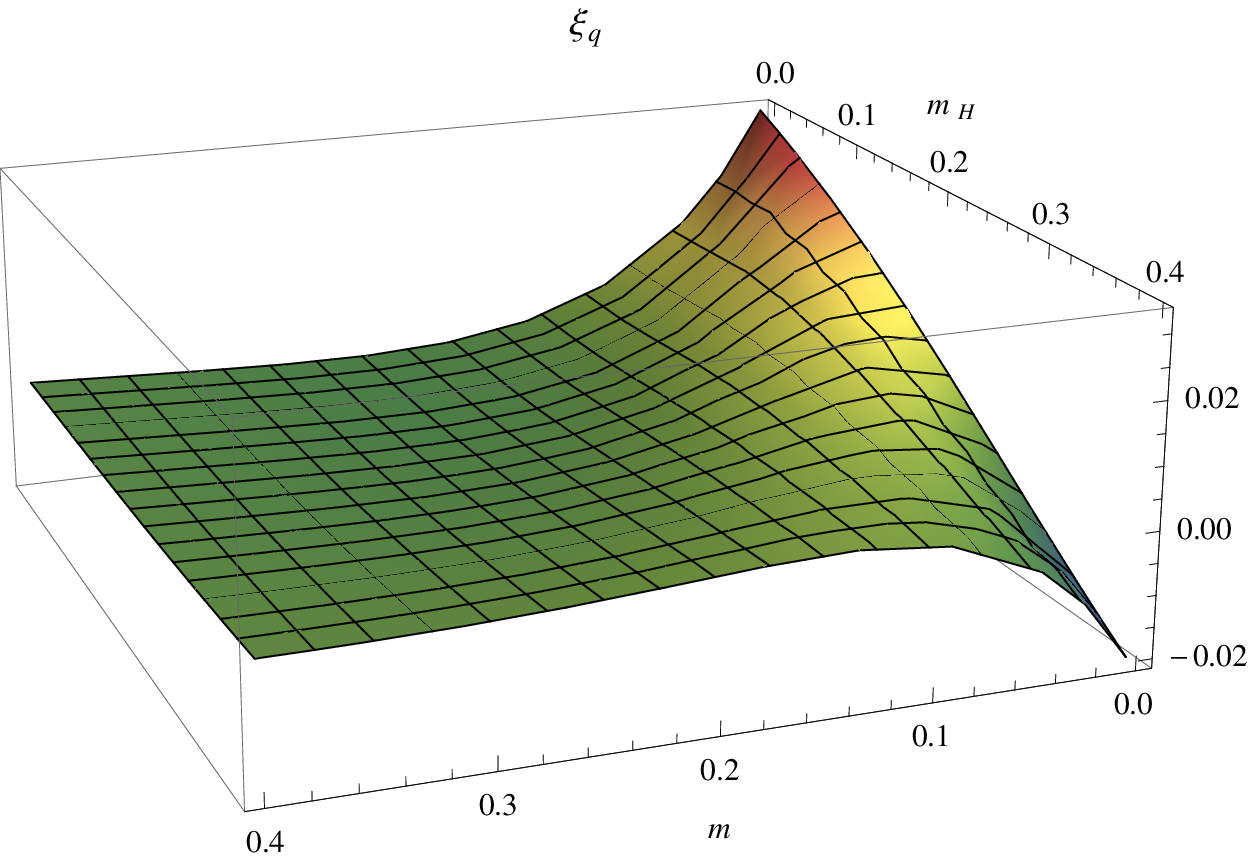}
\includegraphics[width=8.0cm]{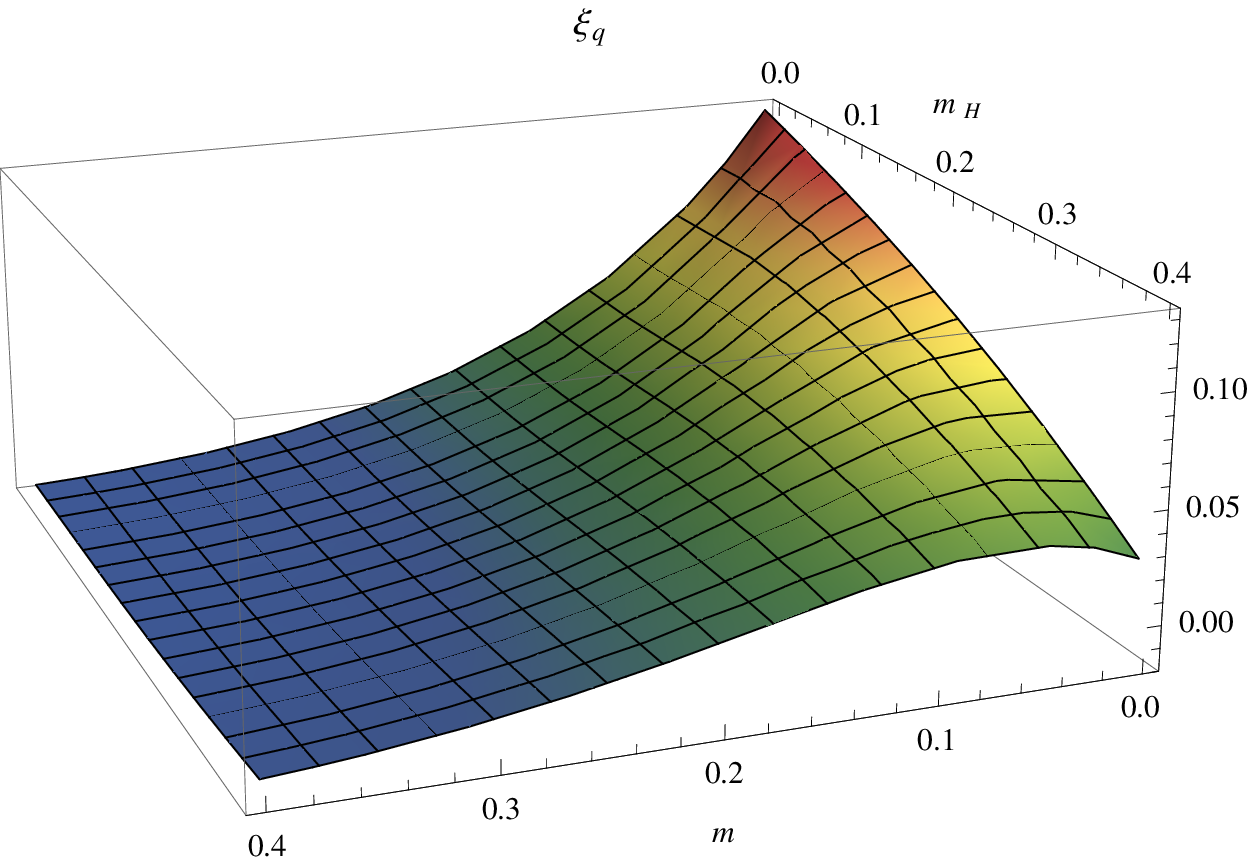}}
\end{center}
\caption{[Color online] Magnitude and sign of the pion-pion interaction in the finite-temperature quark condensate measured by
$\xi_{q}(t,m,m_H)$ -- Eq.~(\ref{xiquark}) -- referring to the temperatures $T= 108 \, MeV$ (left) and $T= 215 \, MeV$ (right).}
\label{figure2}
\end{figure}

To asses the magnitude of the interaction, in Fig.~\ref{figure2}, we plot the dimensionless ratio
\begin{equation}
\label{xiquark}
\xi_{q}(t,m,m_H) = \frac{q_2 T^2}{q_1 F^2}
\end{equation}
that measures the effect of the pion-pion interaction in the quark condensate relative to the free pion gas contribution. The dimensionless
quantities $t, m$, and $m_H$,
\begin{equation}
t = \frac{T}{4 \pi F} \, , \qquad m = \frac{M_{\pi}}{4 \pi F} \, , \qquad m_H = \frac{\sqrt{|qH|}}{4 \pi F} \, , 
\end{equation}
that we use in the figures, capture temperature, pion mass, and strength of the magnetic field relative to the chiral symmetry breaking
scale $\Lambda_{\chi} \approx 4 \pi F \approx \, 1 GeV$. The quantities $t, m$, and $m_H$ must be small since chiral perturbation
theory is a low-energy effective theory. Inspecting Fig.~\ref{figure2} -- where we have chosen $T= 108 \, MeV$ and $T= 215 \, MeV$ as well
as $m, m_H \le 0.4$ -- one notices that the interaction is largest in the chiral limit ($m \to 0$) when no magnetic field is present or
when the magnetic field becomes stronger. The effect of the interaction is not tiny -- rather it may constitute up to about ten percent
relative to the leading free pion gas contribution.

\begin{figure}
\begin{center}
\hbox{
\includegraphics[width=8.0cm]{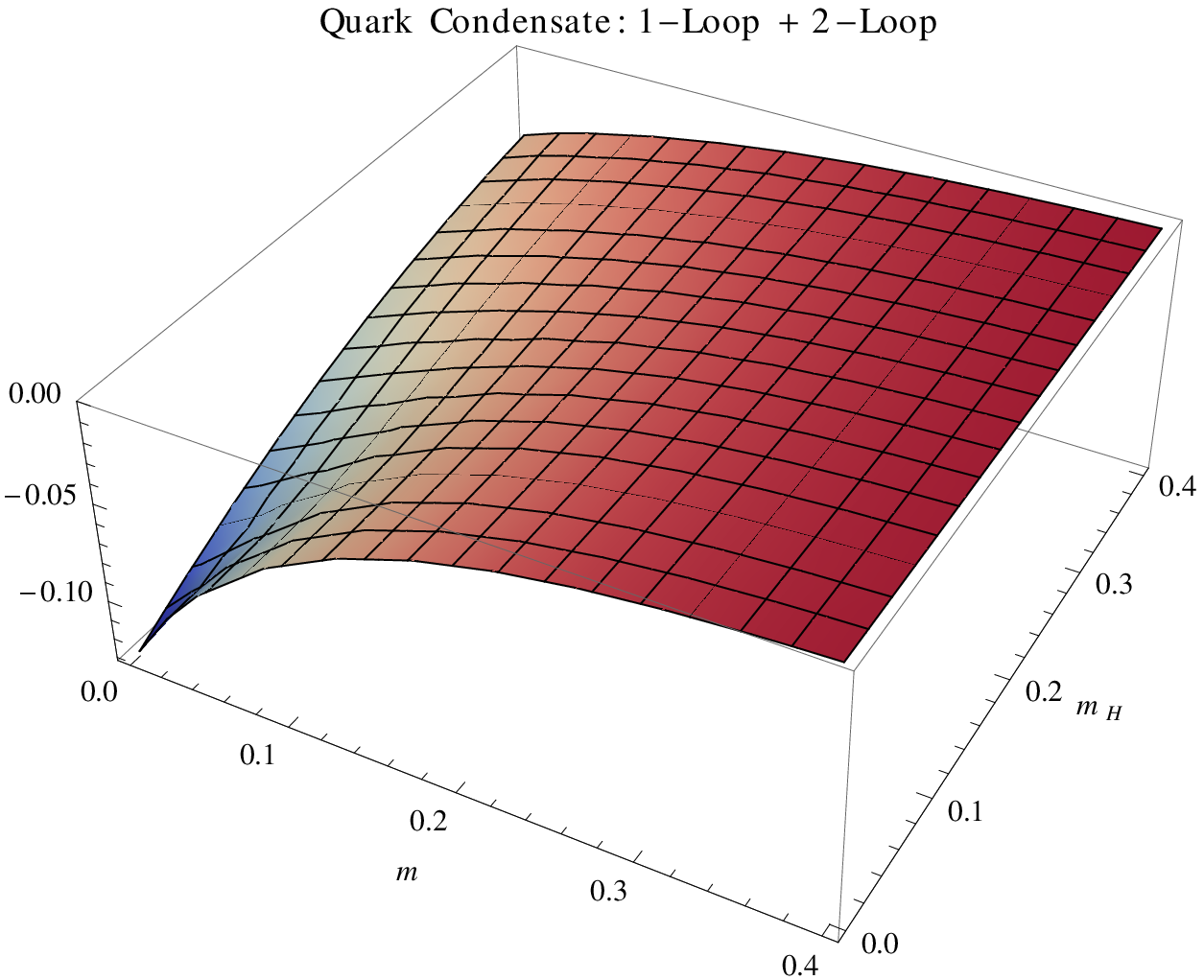}
\includegraphics[width=8.0cm]{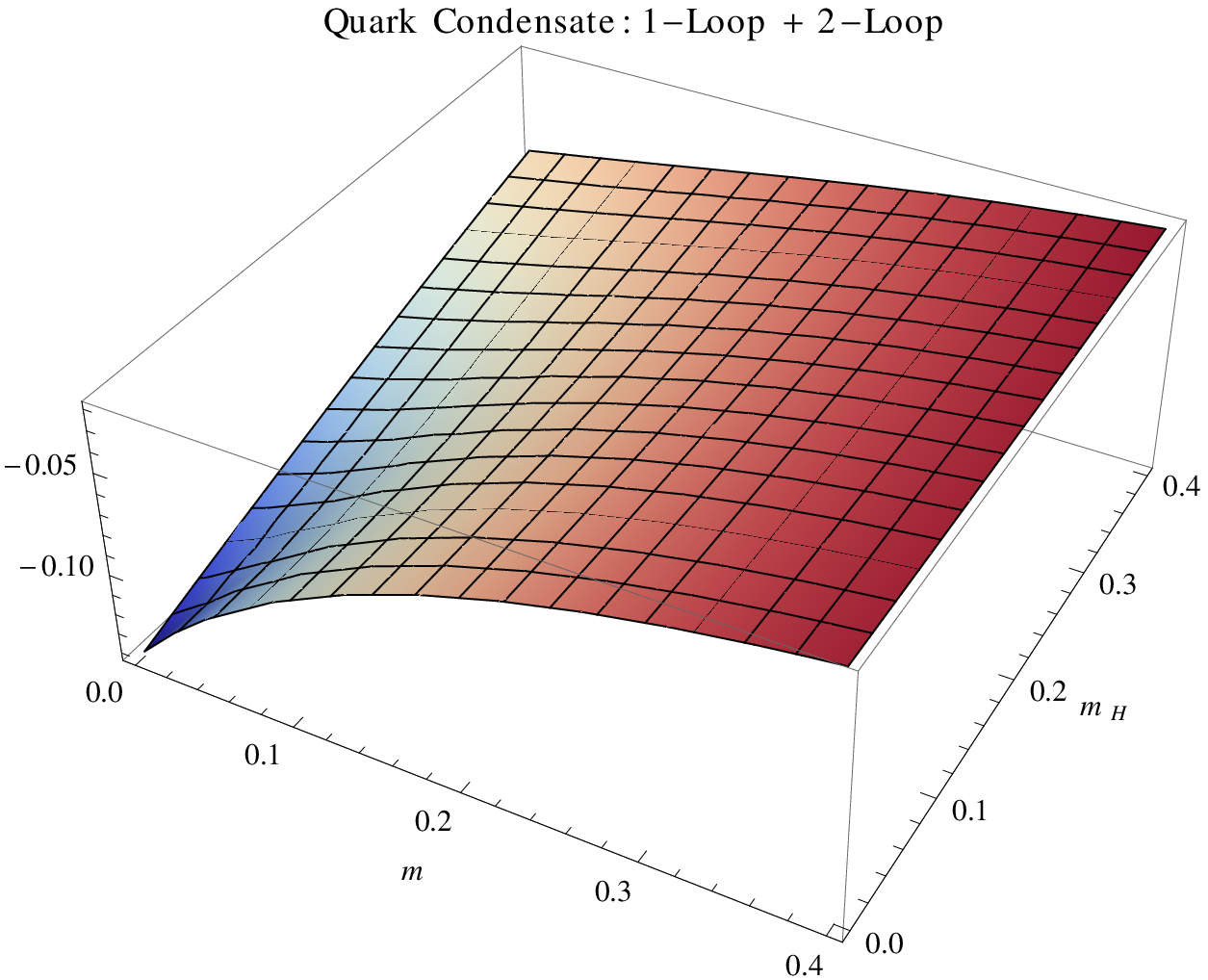}}
\end{center}
\caption{[Color online] Finite-temperature quark condensate: Sum of one- and two-loop contribution at $T= 108 \, MeV$ (left) and
$T= 215 \, MeV$ (right).}
\label{figure3}
\end{figure}

In Fig.~\ref{figure3}, we depict the sum of one- and two-loop contribution, i.e., the dimensionless quantity
\begin{equation}
- \Big( q_1 + q_2 \frac{T^2}{F^2} \Big) \, ,
\end{equation}
for the same two temperatures $T= \{ 108 \, MeV, 215 \, MeV \}$, or, $t= \{ 0.1, 0.2 \}$. As the plots indicate -- at fixed $M_{\pi}$ and
temperature -- the finite-temperature quark condensate increases when the magnetic field grows. The effect is most pronounced in the chiral
limit ($m \to 0$).

\begin{figure}
\begin{center}
\hbox{
\includegraphics[width=8.0cm]{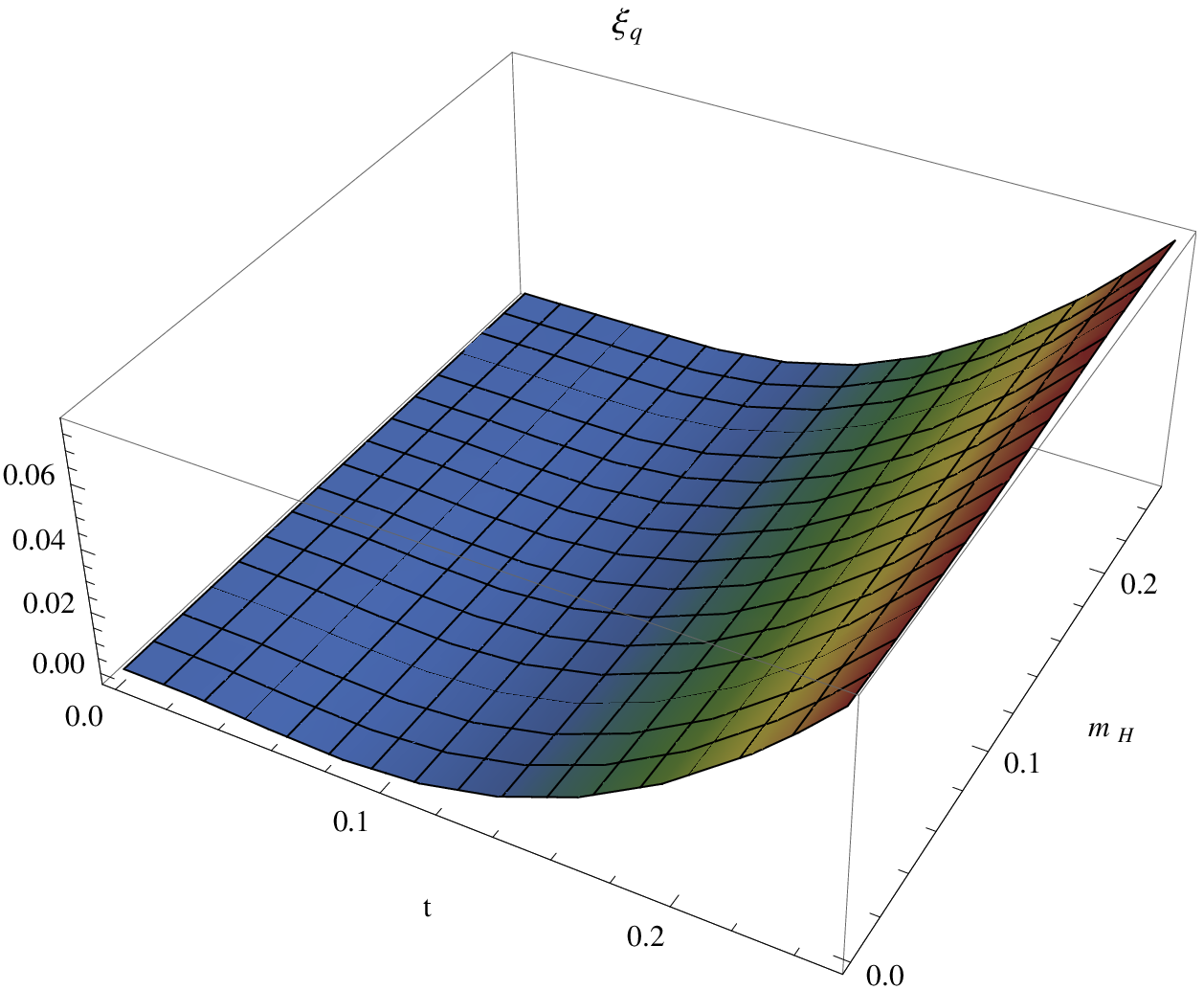}
\includegraphics[width=8.0cm]{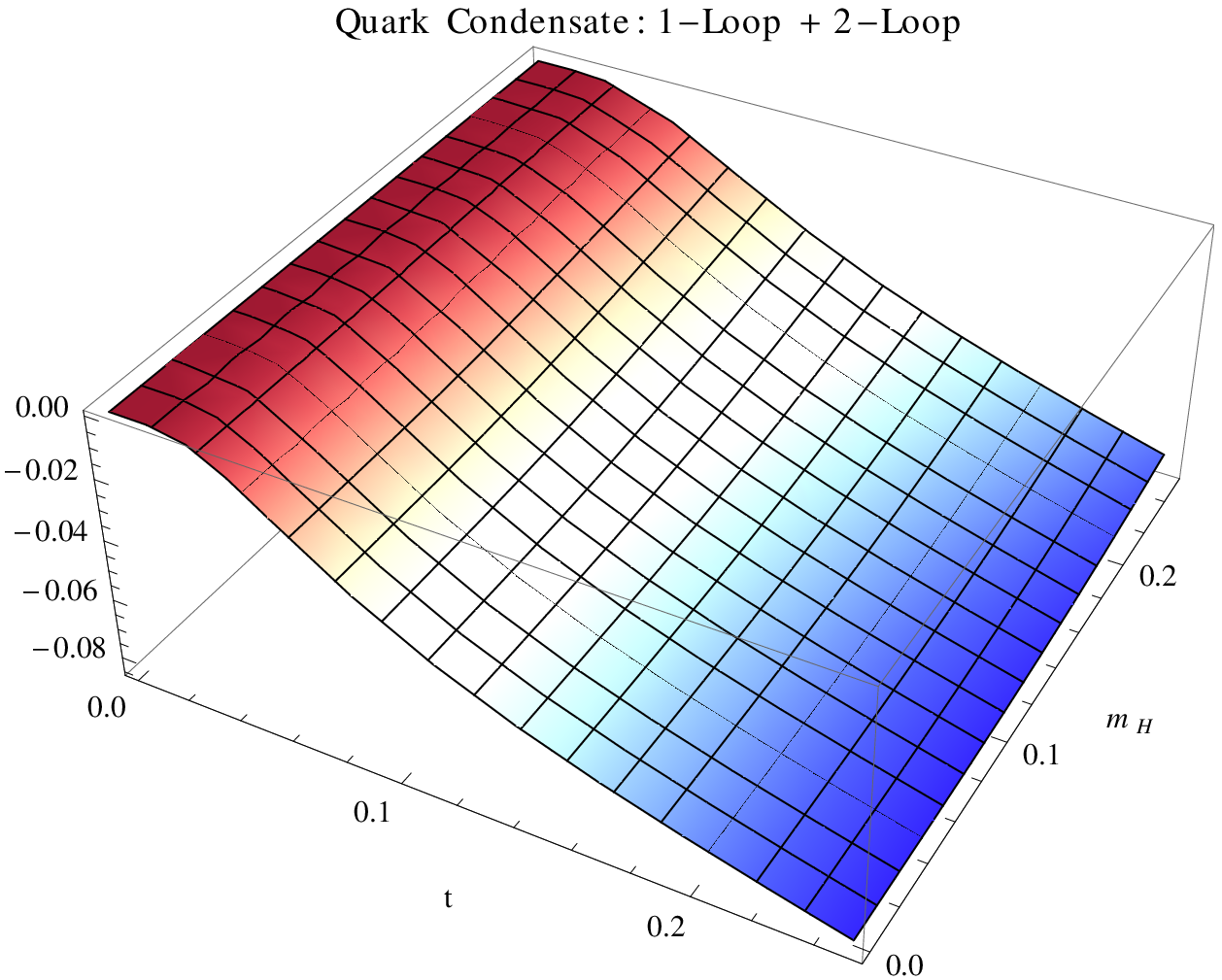}}
\end{center}
\caption{[Color online] LHS: Magnitude and sign of the pion-pion interaction in the finite-temperature QCD quark condensate measured by
$\xi_{q}(t,m,m_H)$ at the physical value $M_{\pi} = 140 \, MeV$. RHS: Sum of one- and two-loop contribution of the finite-temperature quark
condensate at the physical value $M_{\pi}$.}
\label{figure4}
\end{figure}

Let us examine the real world, where the pion masses are fixed at their physical values $M_{\pi} = 140 \, MeV$ ($m = 0.130$)\footnote{For
the tree-level pion decay constant we use the value $F = 85.6 \, MeV$ reported in Ref.~\citep{Aoki20}. Note that in the isospin limit --
and in zero magnetic field -- the masses of the neutral and the charged pions are identical.}. In Fig.~\ref{figure4}, on the LHS, we plot
the ratio $\xi_{q}$ as a function of temperature and magnetic field strength. The effect of the pion-pion interaction is less than ten
percent in the parameter range $t, m_H \le 0.25$ ($T \le 269 \, MeV, \sqrt{|qH|} \le 269 \, MeV$) we are considering. Finally, on the RHS
of Fig.~\ref{figure4}, we depict the sum of one- and two-loop contribution in the quark condensate for the same parameter domain. One
observes that the finite-temperature quark condensate slightly increases when the strength of the magnetic field grows while temperature is
held constant. This effect however is small.

In the chiral limit, the finite-temperature quark condensate reduces to
\begin{eqnarray}
\label{quarkCondensateTChiralLimit}
\frac{{\langle {\bar q} q \rangle}^T}{{\langle 0 | {\bar q} q | 0 \rangle}_0} & = &
- \frac{T^2}{F^2} \, \Bigg\{ \frac{1}{24} + h_1(M_H,T,0) - \Big( \frac{|qH| \ln 2}{32 \pi^2 F^2} \Big) \, h_1(0,T,0)
+ {\tilde h_1}(M_H,T,H) \Bigg\} \nonumber \\
& & + \frac{T^4}{24 F^4} \, \Bigg\{ \frac{1}{48} -  h_1(M_H,T,0) - {\tilde h}_1(M_H,T,H) \Bigg\} + {\cal O}( T^6) \, .
\end{eqnarray}
The mass $M_H$ depends on the magnetic field, 
\begin{equation}
M^2_H = \frac{{\overline l}_6 - {\overline l}_5}{48 \pi^2} \, \frac{{|qH|}^2}{F^2} \, ,
\end{equation}
and corresponds to the charged pion mass in the chiral limit. The mass of the neutral pion, on the other hand, tends to zero in the chiral
limit\footnote{See Eqs.~(\ref{chargedPionMass}) and (\ref{neutralPionMass}).}.

We now address the question of how the quark condensate in the chiral limit behaves in weak magnetic fields. In this limit -- implemented
by $|qH| \ll T^2$ -- we have to expand the kinematical functions $h_1(M_H,T,0)$ and ${\tilde h}_1(M_H,T,H)$ in
Eq.~(\ref{quarkCondensateTChiralLimit}) in the magnetic-field dependent mass $M_H$, which leads to
\begin{eqnarray}
\label{expansionMH}
h_1(M_H,T,0) & = & h_1(0,T,0) - \alpha \epsilon^2 h_2(0,T,0) + \frac{\alpha^2 \epsilon^4}{2!} \, h_3(0,T,0) + {\cal O}(h_4) \, , \nonumber \\
{\tilde h}_1(M_H,T,H) & = & {\tilde h}_1(0,T,H) - \alpha \epsilon^2 {\tilde h}_2(0,T,H)
+ \frac{\alpha^2 \epsilon^4}{2!} \, {\tilde h}_3(0,T,H) + {\cal O}( {\tilde h}_4) \, ,
\end{eqnarray}
with 
\begin{equation}
\alpha = \frac{{\overline l}_6 - {\overline l}_5}{3} \, t^2 \, , \qquad t = \frac{T}{4 \pi F} \, .
\end{equation}
The structure of this infinite series of kinematical functions is analyzed in Appendix \ref{appendixC}. While the functions $h_1(0,T,0)$
and ${\tilde h}_1(0,T,H)$ are well-defined, it should be pointed out that for $r=2,3,4, \dots$, the
functions $h_r(0,T,0)$ and ${\tilde h}_r(0,T,H)$ generate various types of divergences in the weak magnetic field expansion parameter
$\epsilon$. The notation $h_2(0,T,0), {\tilde h}_2(0,T,H), h_3(0,T,0), {\tilde h}_3(0,T,H), \dots$ is therefore symbolic: it is understood
that these functions contain inverse powers of $\epsilon$ as well as logarithms $\ln \epsilon$. These pieces -- according to
Eq.~(\ref{expansionMH}) -- are then multiplied by even powers of $\epsilon$, in such a way that all divergences ultimately disappear in the
quark condensate, as we show in Appendix \ref{appendixC}. The outcome is the following series for the finite-temperature two-loop quark
condensate in the chiral limit and in weak magnetic fields:
\begin{eqnarray}
\label{condensateMySeries}
\frac{{\langle {\bar q} q \rangle}^T}{{\langle 0 | {\bar q} q | 0 \rangle}_0} & = & - \frac{1}{8 F^2} \, T^2 
+ \frac{1}{F^2} \, \Bigg\{ \frac{|I_{\frac{1}{2}}|}{8 \pi^{3/2}} \, \sqrt{\epsilon}
-\frac{\ln 2}{16 \pi^2} \, \epsilon \nonumber \\
& & - \frac{\sqrt{2} -4}{8} \, \gamma \, \zeta(\mbox{$ \frac{3}{2}$})\, \epsilon^{3/2}
+ \frac{\gamma}{4 \pi} \, \epsilon^2 \ln \epsilon
+ {\cal O}(\epsilon^2) \Bigg\} \, T^2 \nonumber \\
& & - \frac{1}{384 F^4} \, T^4 + \frac{1}{F^4} \, \Bigg\{ \frac{|I_{\frac{1}{2}}|}{192 \pi^{3/2}} \, \sqrt{\epsilon} \nonumber \\
& & - \frac{\sqrt{2} -4}{192} \, \gamma \, \zeta(\mbox{$ \frac{3}{2}$}) \, \epsilon^{3/2}
+ \frac{\gamma}{96 \pi} \, \epsilon^2 \ln \epsilon
+ {\cal O}(\epsilon^2) \Bigg\} \, T^4 + {\cal O}(T^6) \, .
\end{eqnarray}
Recall that $\epsilon$,
\begin{equation}
\epsilon = \frac{|qH|}{T^2} \, ,
\end{equation}
is the relevant expansion parameter, while the other quantities are
\begin{eqnarray}
I_{\frac{1}{2}} & = & {\int}_{\!\!\! 0}^{\infty} \, d\rho \rho^{-1/2} \Big( \frac{1}{\sinh(\rho)} - \frac{1}{\rho} \Big) \approx -1.516256 \, ,
\nonumber \\
\gamma & = & \frac{{\overline l}_6 - {\overline l}_5}{12 \pi} \, t^2 \, , \qquad t = \frac{T}{4 \pi F} \, .
\end{eqnarray}
The first two lines of Eq.~(\ref{condensateMySeries}) refer to one-loop order ($\propto T^2$), while the remaining two lines represent
two-loop corrections ($\propto T^4$). In the chiral limit, the series for the finite-temperature quark condensate in weak magnetic fields
is thus characterized by square-root terms $\propto \sqrt{\epsilon}$, a term linear in $\epsilon$, followed by half-integer powers
$\epsilon^{3/2}$ and logarithmic contributions of the form $\epsilon^2 \ln \epsilon$. The remaining contributions involve even powers of
$\epsilon$. Notice that the leading corrections -- proportional to $\sqrt{\epsilon}$ -- come with a positive sign: in the chiral limit, as
already illustrated by Fig.~\ref{figure3}, the finite-temperature quark condensate grows if the magnetic field is switched on.

The published results in Refs.~\citep{Aga00,Aga01,AS01,Aga08,And12a,And12b} do not quite agree with the above representation. The correct
series at one-loop order has been derived and discussed in Ref.~\citep{Hof19}. The two-loop contribution in nonzero magnetic fields,
displayed in the second brace of Eq.~(\ref{condensateMySeries}), again differs from the published two-loop result, Eq.(5.8) of
Ref.~\citep{And12b}: the term
\begin{equation}
\frac{5 \sqrt{|qH|} T^3}{1536 \pi F^4}
\end{equation}
in Eq.(5.8) of Ref.~\citep{And12b} should rather read
\begin{equation}
\frac{\sqrt{|qH|} T^3}{192 \pi^{3/2} F^4} \, |I_{\frac{1}{2}}| \, .
\end{equation}
The numerical discrepancy is
\begin{equation}
\frac{5}{1536 \pi} \approx 0.00103616 \, , \qquad \frac{|I_{\frac{1}{2}}|}{192 \pi^{3/2}} \approx 0.00141823 \, .
\end{equation}
Moreover, a term linear in $\epsilon$ in the second brace of Eq.~(\ref{condensateMySeries}) does not emerge in our expansion --
contradicting the result announced in Ref.~\citep{And12b}. It should be emphasized that the series provided in the literature is restricted
to linear order in $\epsilon$, while we have analyzed the full structure of the weak magnetic field expansion of the finite-temperature
quark condensate in the chiral limit up to two loops.

\subsection{Zero-Temperature Quark Condensate}

We now turn to the quark condensate at zero temperature:
\begin{equation}
\langle 0 | {\bar q} q | 0 \rangle = -\frac{{\langle 0 | {\bar q} q | 0 \rangle}_0}{F^2} \, \frac{\mbox{d} z_0}{\mbox{d} M^2} \, .
\end{equation}
Recall that ${\langle 0 | {\bar q} q | 0 \rangle}_0$ is the quark condensate at $T$=0, $H$=0 and $M$=0. On the basis of the representations
Eqs.~(\ref{freeEDp6ZeroT}) and (\ref{freeEDp4ZeroT}) for the vacuum energy density, we derive
\begin{eqnarray}
\label{quarkCondensatep6ZeroT}
\frac{\langle 0 | {\bar q} q | 0 \rangle}{{\langle 0 | {\bar q} q | 0 \rangle}_0} & = &
1 - \frac{{\overline l}_3 - 4{\overline h}_1}{32 \pi^2} \, \frac{M^2}{F^2}
- \frac{K_1}{F^2}
+ \frac{3 {\overline l}_3}{1024 \pi^4} \, \frac{M^4}{F^4}
- \frac{9 {\overline l}_3 ({\overline c}_{10}+2{\overline c}_{11})}{1024 \pi^4} \, \frac{M^4}{F^4} \nonumber \\
& & - \frac{{\overline l}_6 - {\overline l}_5}{768 \pi^4} \,  \frac{{|qH|}^2}{F^4}
+ \frac{({\overline l}_6 - {\overline l}_5) {\overline c}_{34}}{768 \pi^4} \, \frac{{|qH|}^2}{F^4}
- \frac{1}{32 \pi^2} \, \frac{M^2}{F^4} \, K_1 \nonumber \\
& & + \frac{{\overline l}_3}{16 \pi^2} \, \frac{M^2}{F^4} \, K_1
+ \frac{{\overline l}_3}{32 \pi^2} \, \frac{M^4}{F^4} \, \frac{\mbox{d} K_1}{ \mbox{d} M^2}
- \frac{({\overline l}_6 - {\overline l}_5)}{48 \pi^2} \, \frac{{|qH|}^2}{F^4} \, \frac{\mbox{d} K_1}{ \mbox{d} M^2} \, .
\end{eqnarray}
The explicit expressions for $K_1$ and $\mbox{d} K_1/\mbox{d} M^2$,
\begin{eqnarray}
K_1 & = & \frac{M^2}{16 \pi^2} - \frac{M^2}{16 \pi^2} \, \ln \frac{M^2}{2|qH|} + \frac{|qH|}{8 \pi^2} \,
\ln \Gamma \! \Big( \frac{M^2}{2|qH|} + \frac{1}{2} \Big) - \frac{|qH|}{16 \pi^2} \, \ln 2\pi \, , \nonumber \\
\frac{\mbox{d} K_1}{\mbox{d} M^2} & = & \frac{1}{16 \pi^2} \, \ln \frac{|qH|}{M^2} + \frac{1}{16 \pi^2} \,
\Psi \! \Big( \frac{M^2}{2|qH|} + \frac{1}{2} \Big) + \frac{\ln 2}{16 \pi^2} \, ,
\end{eqnarray}
are derived in Appendix \ref{appendixB}. The series for the quark condensate is organized according to ascending powers of $M^2$ and $|qH|$
-- both quantities count as order $p^2$. The respective coefficients depend in a nontrivial manner on the ratio $M^2/|qH|$ and involve
renormalized NLO and NNLO effective constants. Let us compare our result with the literature.

The focus of the two-loop  CHPT calculation presented in Ref.~\citep{Wer08}, was to determine the shift in the zero-temperature quark
condensate caused by an external (electro)magnetic field. Our expression, Eq.~(\ref{quarkCondensatep6ZeroT}), goes beyond the literature
since we have derived the whole two-loop representation for the quark condensate -- not just the terms induced by the magnetic field.

To analyze the chiral limit of the zero-temperature quark condensate in finite magnetic fields, we invoke the behavior of the NLO and NNLO
effective constants. According to Appendix \ref{appendixA} we have 
\begin{eqnarray}
{\overline l}_3, {\overline l}_5, {\overline l}_6  & \propto & \ln M^2 \, , \nonumber \\
{\overline c}_{34} & \propto & \ln M^2 \, , \nonumber \\
{\overline c}_{10} + 2 {\overline c}_{11} & \propto & \ln M^2  \, ,
\end{eqnarray}
i.e., the renormalized NLO and NNLO effective constants explode in the limit $M \to 0$. But note that in the quark condensate these
constants are multiplied by powers of $M^2$ such that the chiral limit is in fact unproblematic. While some terms in
Eq.~(\ref{quarkCondensatep6ZeroT}) hence disappear in the chiral limit, only the following two terms,
\begin{equation}
\frac{({\overline l}_6 - {\overline l}_5)}{768 \pi^4} \, {\overline c}_{34} \, \frac{{|qH|}^2}{F^4}
- \frac{({\overline l}_6 - {\overline l}_5)}{768 \pi^4} \, \frac{{|qH|}^2}{F^4} 
\ln\Big( \frac{|qH|}{M^2} \Big) \, ,
\end{equation}
need special consideration, as they both explode in the chiral limit. However, writing the NNLO effective constant ${\overline c}_{34}$ as
\begin{equation}
{\overline c}_{34} = \ln\Big( \frac{\Lambda^2_{34}}{M^2_{\pi}} \Big) \, ,
\end{equation}
where $\Lambda_{34}$ is the renormalization group invariant scale associated with ${\overline c}_{34}$, the two terms can be merged such
that the zero-temperature quark condensate in nonzero magnetic fields is well-defined in the chiral limit, taking the form
\begin{eqnarray}
\label{quarkCondensatep6ZeroTchiralLimit}
\frac{\langle 0 | {\bar q} q | 0 \rangle}{{\langle 0 | {\bar q} q | 0 \rangle}_0} & = &
1 + \frac{\ln 2}{16 \pi^2} \, \frac{|qH|}{F^2} 
- \frac{({\overline l}_6 - {\overline l}_5)}{768 \pi^4} \, \frac{{|qH|}^2}{F^4} 
\ln\Big( \frac{|qH|}{\Lambda^2_{34}} \Big)
- \frac{({\overline l}_6 - {\overline l}_5)}{768 \pi^4} \, \frac{{|qH|}^2}{F^4} \nonumber \\
& & - \frac{({\overline l}_6 - {\overline l}_5)}{768 \pi^4} \, \frac{{|qH|}^2}{F^4} \,
\Bigg( \frac{\Gamma'(\mbox{$ \frac{1}{2}$})}{\Gamma(\mbox{$ \frac{1}{2}$})} + \ln 2 \Bigg) \, .
\end{eqnarray}
Notice that the $\ln M^2$-dependence in the combination ${\overline l}_6 - {\overline l}_5$ cancels, and we can write
\begin{equation}
{\overline l}_6 - {\overline l}_5 = \ln\Big( \frac{\Lambda^2_6}{\Lambda^2_5} \Big) \, ,
\end{equation}
where $\Lambda_5$ and $\Lambda_6$ are the respective renormalization group invariant scales associated with the NLO effective constants
${\overline l}_5$ and ${\overline l}_6$.

\section{Conclusions}
\label{conclusions}

We have explored the behavior of the quark condensate subjected to an external magnetic field within the framework of chiral perturbation
theory. Unlike previous two-loop evaluations by other authors, we have used a coordinate space representation.

Regarding the finite-temperature quark condensate in the chiral limit and in weak magnetic fields, we have pointed out various errors that
have occurred in the literature and have provided the correct series. At order $T^2$ -- and in terms of the expansion parameter
$\epsilon = |qH|/T^2$ -- the leading contribution is proportional to $\sqrt{\epsilon}$, followed by a term linear in $\epsilon$, a
half-integer power $\epsilon^{3/2}$ and a logarithmic contribution $\epsilon^2 \ln \epsilon$. The remaining contributions involve even
powers of $\epsilon$. At order $T^4$ the pattern repeats itself with the exception that a term linear in $\epsilon$ does not occur.

Leaving the weak magnetic field limit, we have investigated the impact of the magnetic field on the quark condensate at finite temperature.
Emphasis was put on the effect of the pion-pion interaction which constitutes up to about ten percent for arbitrary pion masses but also in
the real world where $M_{\pi} = 140 \, MeV$. The interaction is largest in the chiral limit. The finite-temperature quark condensate (sum of
one- and two-loop contribution) at fixed temperature and fixed pion mass grows monotonically when the magnetic field strength increases.
Again, the effect is most pronounced in the chiral limit.

Finally we have derived the two-loop representation for the QCD vacuum energy density and the quark condensate at zero temperature. We have
complemented earlier studies by other authors, by providing the full two-loop representation, i.e., not just the terms that emerge on
account of the nonzero magnetic field.

A natural -- but highly nontrivial -- step is to extend the present analysis to the three-loop level, in analogy to the three-loop analysis
in zero magnetic field given in the pioneering article \citep{GL89}, based on a coordinate space representation of CHPT. Corresponding work
is in progress.

\section*{Acknowledgments}
The author gratefully acknowledges H.\ Leutwyler and J.\ Bijnens for correspondence. Special thanks to J.\ Bijnens for sharing unpublished
results on the order-$p^6$ zero-temperature quark condensate.

\begin{appendix}

\section{Order-$p^6$ Free Energy Density at $T$=0}
\label{appendixA}

\subsection{Low-Energy Effective Constants at NLO and NNLO}
\label{appendixA1}

The aim of the present appendix is to discuss the renormalization group running of the NLO and NNLO effective constants $l^r_i$ and $c^r_i$
in some detail, and then to provide a definition of the renormalized NNLO effective constants ${\overline c}_i$ -- in analogy to the
definition of the renormalized NLO quantities ${\overline l}_i$.

The NNLO effective constants $c_i$ that appear in ${\cal L}^6_{eff}$, are defined in Ref.~\citep{BCE00} as
\begin{equation}
\label{definitionci}
c_i = \frac{{(c \mu)}^{2(d-4)}}{F^2} \, \Bigg\{ c^r_i - \gamma^{(2)}_i \Lambda^2 - \gamma^{(1)}_i \Lambda - \gamma^{(L)}_i \Lambda \Bigg\} \, ,
\end{equation}
with
\begin{equation}
\Lambda = \frac{1}{16 \pi^2} \, \frac{1}{d-4} \, , \qquad \ln c = -\mbox{$ \frac{1}{2}$} \Big[ \ln 4 \pi + \Gamma'(1) +1 \Big] \, .
\end{equation}
The quantities $\gamma^{(1)}_i, \gamma^{(2)}_i$ are pure numbers and the $c^r_i$ are the renormalized running NNLO effective constants. For
the definition of the NLO effective constants $l_i$ that appear in ${\cal L}^4_{eff}$, on the other hand, we adopt the definition given in
the original Ref.~\citep{GL84},
\begin{equation}
\label{definitionli}
l_i = l^r_i + \gamma_i \lambda \, ,
\end{equation}
where 
\begin{eqnarray}
\lambda & = & \mbox{$ \frac{1}{2}$} \, (4 \pi)^{-\frac{d}{2}} \, \Gamma(1-{\mbox{$ \frac{1}{2}$}}d) \mu^{d-4} \nonumber \\
& & = \frac{\mu^{d-4}}{16{\pi}^2} \, \Bigg[ \frac{1}{d-4} - \mbox{$ \frac{1}{2}$} \{ \ln{4{\pi}} + {\Gamma}'(1) + 1 \}
+ {\cal O}(d\!-\!4) \Bigg] \, .
\end{eqnarray}
The $\gamma_i$ are pure numbers and the $l^r_i$ are the renormalized running NLO effective constants. The definition,
Eq.~(\ref{definitionli}), can be rewritten as
\begin{equation}
l_i = l^r_i + {(c \mu)}^{d-4} \gamma_i \Lambda \, .
\end{equation}
Note that the $\gamma_i$ also show up in $\gamma^{(L)}_i$, Eq.~(\ref{definitionci}), in the form of
\begin{equation}
\gamma^{(L)}_i = \sum_j \gamma^{(L)}_{ij} {(c \mu)}^{-(d-4)} l^r_j \, ,
\end{equation}
where the coefficients $\gamma^{(L)}_{ij}$ are again pure numbers.

Since the $c_i$ do not depend on the renormalization scale $\mu$, one concludes that the renormalization group running of the NNLO
effective constants $c^r_i$ is
\begin{equation}
\label{relationcr}
\mu \frac{\mbox{d} c^r_{i}}{\mbox{d} \mu} = - 2(d-4) c^r_i + \frac{\gamma^{(1)}_i}{8 \pi^2} + \frac{\gamma^{(L)}_i}{16 \pi^2} \, .
\end{equation}
In the above derivation we have used the fact that the NLO effective constants $l^r_i$ themselves obey the running
\begin{equation}
\label{relationlr}
\mu \frac{\mbox{d} l^r_i}{\mbox{d} \mu} = - \frac{\gamma_i}{16 \pi^2} \, {(c \mu)}^{d-4} \, ,
\end{equation}
which follows from the fact that the $l_i$ do not depend on $\mu$. Furthermore, with the Weinberg consistency condition \citep{Wei79},
\begin{equation}
-2 \gamma^{(2)}_i + \sum_j \gamma^{(L)}_{ij} \gamma_j = 0 \, ,
\end{equation}
a divergence linear in $\Lambda$ has been eliminated in Eq.~(\ref{relationcr}).

Instead of the NLO quantities $l^r_i$ that depend on the renormalization scale, alternatively one often uses the NLO effective constants
${\overline l}_i$ that are $\mu$-independent. The connection between the two is \citep{GL84}
\begin{equation}
\label{definitionliBar}
l^r_i = \frac{\gamma_i}{32 \pi^2} \Big( {\overline l}_i + \ln \frac{M^2}{\mu^2} \Big) \, .
\end{equation}
Let us transfer this connection to NNLO. The specific NNLO effective constants that appear in the vacuum energy density are $c_{10}, c_{11}$
and $c_{34}$, where the last one only matters when a magnetic field is present. Following Ref.~\citep{BCE00} -- but using the convention
(\ref{definitionli}) -- it reads
\begin{equation}
c_{34} = \frac{{(c \mu)}^{2(d-4)}}{F^2} \, c^r_{34} + \frac{l^r_5 - \mbox{$ \frac{1}{2}$} l^r_6}{F^2} \, \lambda \, .
\end{equation}
Explicitly, the running of $c^r_{34}$ is given by
\begin{equation}
\frac{\mbox{d} c^r_{34}}{\mbox{d} \mu^2} = - \frac{l^r_5 - \mbox{$ \frac{1}{2}$} l^r_6}{32 \pi^2 \mu^2} \, .
\end{equation}
In analogy to the above definition for the NLO constants ${\overline l}_i$, Eq.~(\ref{definitionliBar}), that is based on the running
(\ref{relationlr}), we define the renormalized NNLO effective constant ${\overline c}_{34}$ as
\begin{equation}
\label{c34Definition}
c^r_{34} = \frac{{\overline l}_6 - {\overline l}_5}{6144 \pi^4} \, {\overline c}_{34}
+ \frac{{\overline l}_6 - {\overline l}_5}{6144 \pi^4} \, \ln \frac{M^2}{\mu^2} \, .
\end{equation}
Note that we have used
\begin{equation}
\gamma_5 = - \frac{1}{6} \, , \qquad \gamma_6 = - \frac{1}{3} \, .
\end{equation}
Since $c^r_{34}$ does not depend on $M$, we conclude
\begin{equation}
\frac{\mbox{d} {\overline c}_{34}}{\mbox{d} M^2} = - \frac{1}{M^2} \, .
\end{equation}
The NNLO constant ${\overline c}_{34}$ hence obeys the same simple relation as the NLO constants ${\overline l}_i$,
\begin{equation}
\frac{\mbox{d} {\overline l}_i}{\mbox{d} M^2} = - \frac{1}{M^2} \, .
\end{equation}

Next we consider the NNLO effective constants $c_{10}$ and $c_{11}$ that arise in the tree-level contribution $z_{6C}$ in the absence of the
magnetic field. They are defined as (see Ref.~\citep{BCE00})
\begin{eqnarray}
c_{10} & = & \frac{{(c \mu)}^{2(d-4)}}{F^2} \, c^r_{10} + \frac{3}{64 F^2}\, \lambda^2
-\frac{1}{F^2} \Big( \mbox{$ \frac{3}{16}$} l^r_3 + \mbox{$ \frac{1}{16}$} l^r_7 \Big) \lambda \, , \nonumber \\
c_{11} & = & \frac{{(c \mu)}^{2(d-4)}}{F^2} \,  c^r_{11} - \frac{9}{128 F^2} \, \lambda^2
+ \frac{1}{F^2} \, \Big( \mbox{$ \frac{9}{32}$} l^r_3 + \mbox{$ \frac{1}{32}$} l^r_7 \Big) \lambda \, .
\end{eqnarray}
Note that in the linear combination $c_{10} + 2c_{11}$ -- as it appears in the vacuum energy density at order $p^6$ -- the dependence on
$l^r_7$ cancels and we are left with
\begin{equation}
c_{10} + 2c_{11} = \frac{{(c \mu)}^{2(d-4)}}{F^2} \,  (c^r_{10} + 2 c^r_{11}) - \frac{3}{32 F^2}\, \lambda^2
+ \frac{3}{8 F^2} \, l^r_3 \lambda \, .
\end{equation}
Since the $c_i$ do not depend on the renormalization scale $\mu$, we conclude
\begin{equation}
\frac{\mbox{d} ( c^r_{10} + 2c^r_{11} )}{\mbox{d} \mu^2} = - \frac{3 l^r_3}{256 \pi^2} \, \frac{1}{\mu^2}  \, .
\end{equation}
Equivalently, by making the replacement $l^r_3 \to {\overline l}_3$,
\begin{equation}
l^r_3 = \frac{\gamma_3}{32 \pi^2} \Big( {\overline l}_3 + \ln \frac{M^2}{\mu^2} \Big) \, , \qquad \gamma_3 = -\frac{1}{2} \, ,
\end{equation}
we can write
\begin{equation}
\frac{\mbox{d} ( c^r_{10} + 2c^r_{11} )}{\mbox{d} \mu^2} = \frac{3 {\overline l}_3}{16384 \pi^4} \, \frac{1}{\mu^2}
+ \frac{3}{16384 \pi^4} \, \frac{1}{\mu^2} \, \ln \frac{M^2}{\mu^2} \, .
\end{equation}
This leads us to the definition of the renormalized combination ${\overline c}_{10} + 2{\overline c}_{11}$ as
\begin{equation}
\label{c10c11Definition}
c^r_{10} + 2 c^r_{11} = - \frac{3 {\overline l}_3}{16384 \pi^4} \, ( {\overline c}_{10} + 2 {\overline c}_{11} )
- \frac{3 {\overline l}_3}{16384 \pi^4} \, \ln \frac{M^2}{\mu^2}
- \frac{3}{32768 \pi^4} \, {\Big(\ln \frac{M^2}{\mu^2}\Big)}^2 \, .
\end{equation}
By construction, the linear combination ${\overline c}_{10} + 2{\overline c}_{11}$ is independent of $\mu$, much like ${\overline c}_{34}$
and the ${\overline l}_i$. Because the expression $c^r_{10} + 2 c^r_{11}$ does not depend on $M$, we also conclude
\begin{equation}
\frac{\mbox{d} ({\overline c}_{10} + 2{\overline c}_{11})}{\mbox{d} M^2} = - \frac{1}{M^2} + \frac{1}{M^2} \,
\frac{{\overline c}_{10} + 2{\overline c}_{11}}{{\overline l}_3} \, .
\end{equation}

\subsection{Isolating UV-divergences}
\label{appendixA2}

Here we focus on the zero-temperature contributions in the free energy density that emerge at order $p^6$ due to the three diagrams
$6A$-$C$ displayed Fig.~\ref{figure1}. The unrenormalized expressions that contain both $T$=0 and finite-temperature pieces are
\begin{eqnarray}
z_{6A} & = & \frac{M^2}{2 F^2} \, G^{\pm}_1 G^0_1 - \frac{M^2}{8 F^2} \, G^0_1 G^0_1 \, , \nonumber \\
z_{6B} & = & (4l_5 - 2l_6) \frac{{|qH|}^2}{F^2} \, G^{\pm}_1 + 2 l_3 \frac{M^4}{F^2} \, G^{\pm}_1 + l_3 \frac{M^4}{F^2} \, G^0_1  \, ,
\nonumber \\
z_{6C} & = & -16(c_{10} + 2c_{11}) M^6 - 8 c_{34} {|qH|}^2 M^2 \, ,
\end{eqnarray}
where $G^{\pm}_1$ and $G^0_1$ are the thermal pion propagators evaluated at the coordinate origin $x$=0,
\begin{equation}
G^{\pm}_1 = G^{\pm}(0) \, , \qquad G^0_1 = G^0(0) \, .
\end{equation}
Inserting the decomposition of thermal propagators into zero-temperature and finite-temperature pieces (defined in
Eq.~(\ref{boseFunctions}))
\begin{eqnarray}
& & G^{\pm}_1 = \Delta^{\pm}(0) + {\tilde g}_1(M,T,H) + g_1(M,T,0) \, , \nonumber \\
& & G^0_1 = \Delta^0(0) + g_1(M,T,0) \, ,
\end{eqnarray}
and using the representations of the zero-temperature propagators $\Delta^{\pm}(0)$ and $\Delta^0(0)$,
\begin{equation}
\Delta^{\pm}(0) = 2 M^2 \lambda + K_1 \, , \qquad \Delta^0(0) = 2 M^2 \lambda\, ,
\end{equation}
with $K_1$ and $\lambda$ as
\begin{eqnarray}
K_1 & = & \frac{{|qH|}^{\frac{d}{2}-1}}{{(4 \pi)}^{\frac{d}{2}}} \,  \, {\int}_{\!\!\! 0}^{\infty} \mbox{d} \rho \, \rho^{-\frac{d}{2}+1}
\, \exp\Big( -\frac{M^2}{|qH|} \rho \Big) \, \Big( \frac{1}{\sinh(\rho)} - \frac{1}{\rho} \Big) \, , \nonumber \\
\lambda & = & \mbox{$ \frac{1}{2}$} \, (4 \pi)^{-\frac{d}{2}} \, \Gamma(1-{\mbox{$ \frac{1}{2}$}}d) M^{d-4} \nonumber \\
& & = \frac{M^{d-4}}{16{\pi}^2} \, \Bigg[ \frac{1}{d-4} - \mbox{$ \frac{1}{2}$} \{ \ln{4{\pi}} + {\Gamma}'(1) + 1 \}
+ {\cal O}(d\!-\!4) \Bigg] \, ,
\end{eqnarray}
we obtain
\begin{eqnarray}
\label{fedT0NotRenormalized}
z^{0}_{6A} & = & \frac{3 M^6}{2 F^2} \, \lambda^2 + \frac{M^4}{F^2} \, K_1 \lambda \, , \nonumber \\
z^{0}_{6B} & = & 6 l_3 \frac{M^6}{F^2} \, \lambda + 2 l_3 \frac{M^4}{F^2} \, K_1 + (8l_5 - 4l_6) \frac{{M^2 |qH|}^2}{F^2} \, \lambda
+ (4l_5 - 2l_6) \frac{{|qH|}^2}{F^2} \, K_1 \, , \nonumber \\
z^{0}_{6C} & = & -16(c_{10} + 2c_{11}) M^6 - 8 c_{34} {|qH|}^2 M^2 \, .
\end{eqnarray}
The upper index $"0"$ signals that we are considering the $T$=0 part only.\footnote{The finite-temperature contribution $z^T$ is given by
Eq.~(\ref{fedPhysicalM}).} To isolate the UV-divergences in this unrenormalized expression, we use the conventions for the NLO and NNLO
effective constants $l_i$ and $c_i$, respectively, that we have provided in Appendix \ref{appendixA1}. One finds that in the sum of the
three diagrams, all UV-divergences disappear and the renormalized order-$p^6$ vacuum energy density takes the form
\begin{eqnarray}
z^{[6]}_0 & = & z^{0}_{6A} + z^{0}_{6B} + z^{0}_{6C} \nonumber \\
& = & \frac{3{\overline l}_3 ({\overline c}_{10} + 2 {\overline c}_{11})}{1024 \pi^4} \, \frac{M^6}{F^2}
- \frac{({\overline l}_6 - {\overline l}_5) {\overline c}_{34}}{768 \pi^4} \, \frac{{|qH|}^2 M^2}{F^2} \nonumber \\
& & - \frac{{\overline l}_3}{32 \pi^2} \,  \frac{M^4}{F^2} \, K_1
+ \frac{({\overline l}_6 - {\overline l}_5)}{48 \pi^2} \, \frac{{|qH|}^2}{F^2} \, K_1 \, .
\end{eqnarray}
The above representation is renormalization-scale independent. This constitutes a nontrivial check of our calculation.

\section{Analysis of the Integral $K_1$}
\label{appendixB}

To analyze the free energy density and the quark condensate in the chiral limit, we must have a closer look at the dimensionally
regularized integral $K_1$,
\begin{eqnarray}
\label{integralK1}
K_1 & = & \frac{{|qH|}^{\frac{d}{2}-1}}{{(4 \pi)}^{\frac{d}{2}}} \,  \, {\int}_{\!\!\! 0}^{\infty} \mbox{d} \rho \, \rho^{-\frac{d}{2}+1}
\, \exp\Big( -\frac{M^2}{|qH|} \rho \Big) \, \Big( \frac{1}{\sinh(\rho)} - \frac{1}{\rho} \Big) \, .
\end{eqnarray}
To this end we first consider the integral $I_2$, defined in (A1) of Ref.~\citep{Hof19} as
\begin{eqnarray}
I_2 & = & - \frac{{|qH|}^{\frac{d}{2}}}{{(4 \pi)}^{\frac{d}{2}}} {\int}_{\!\!\! 0}^{\infty} d\rho \rho^{-\frac{d}{2}}
\Big( \frac{1}{\sinh(\rho)} - \frac{1}{\rho} \Big) \, \exp\!\Big( -\frac{M^2}{|qH|} \rho \Big) \, , \nonumber \\
& = & - \frac{{|qH|}^3}{96 \pi^2 M^2} - \frac{{|qH|}^{\frac{d}{2}}}{{(4 \pi)}^{\frac{d}{2}}} {\int}_{\!\!\! 0}^{\infty} d\rho \rho^{-\frac{d}{2}}
\Big( \frac{1}{\sinh(\rho)} - \frac{1}{\rho} + \frac{\rho}{6} \Big) \, \exp\!\Big( -\frac{M^2}{|qH|} \rho \Big) \, .
\end{eqnarray}
Comparing these representations, one concludes
\begin{eqnarray}
\label{derivativesI2}
K_1 & = & \frac{\mbox{d} I_2}{\mbox{d} M^2} \, , \nonumber \\
\frac{\mbox{d} K_1}{\mbox{d} M^2} & = & \frac{{\mbox{d}}^2 I_2}{{(\mbox{d} M^2)}^2} \, .
\end{eqnarray}
Using the property of the Riemann zeta function
\begin{equation}
\lim_{s \to 1}  \zeta(s,q) = \frac{1}{s-1} - \frac{\Gamma'(q)}{\Gamma(q)} \, ,
\end{equation}
where
\begin{equation}
\zeta(s,q) = \sum^{\infty}_{n=0} \frac{1}{{(q+n)}^s} \, ,
\end{equation}
the second relation in Eq.~(\ref{derivativesI2}) yields\footnote{The physical limit $d \to 4$ is straightforward and does not pose any
problems.}
\begin{equation}
\frac{\mbox{d} K_1}{\mbox{d} M^2} = \frac{1}{16 \pi^2} \, \ln \frac{|qH|}{M^2} + \frac{1}{16 \pi^2} \,
\Psi \! \Big( \frac{M^2}{2|qH|} + \frac{1}{2} \Big) + \frac{\ln 2}{16 \pi^2} \, ,
\end{equation}
where $\Psi(x)$ is the Polygamma function
\begin{equation}
\Psi(x) = \frac{\Gamma'(x)}{\Gamma(x)} \, .
\end{equation}
The expression for $K_1$ is obtained by integration,
\begin{equation}
K_1 = \frac{M^2}{16 \pi^2} - \frac{M^2}{16 \pi^2} \, \ln \frac{M^2}{2|qH|} + \frac{|qH|}{8 \pi^2} \,
\ln \Gamma \! \Big( \frac{M^2}{2|qH|} + \frac{1}{2} \Big) + C(|qH|) \, .
\end{equation}
The integration constant $C(|qH|)$ can be determined by setting $M$=0 in the equation above and in the original representation,
Eq.~(\ref{integralK1}). One identifies
\begin{equation}
C(|qH|) = - \frac{|qH|}{16 \pi^2}  \ln 2 \pi \, .
\end{equation}
While $K_1$ appears in the free energy density, the derivative $\mbox{d} K_1/\mbox{d} M^2$ is relevant in the quark condensate.

\section{Bose Functions in the Chiral Limit}
\label{appendixC}

The finite-temperature representation of the quark condensate in the chiral limit, Eq.~(\ref{quarkCondensateTChiralLimit}), features an
infinite series of kinematical Bose functions $g_r$ and ${\tilde g}_r$ that has to be resummed because of the weak magnetic field expansion
Eq.~(\ref{expansionMH}). This is the main focus of the present appendix. The aim is to provide explicit expressions up to order
$\epsilon^2 \ln \epsilon $ in the finite-temperature quark condensate.

We first consider the second type of functions\footnote{It should be noted that the functions ${\tilde g_r},g_r$ -- up to temperature
powers -- coincide with the functions ${\tilde h_r},h_r$. The conversion is given by  Eq.~(\ref{conversion}).}
\begin{eqnarray}
\label{ABC}
{\tilde g_r}(M^{\pm}_{\pi}, T, H) & = & \frac{\epsilon}{{(4 \pi)}^{r+1}} T^{d-2r} \, {\int}_{\!\!\! 0}^{\infty} \mbox{d} \rho \, \rho^{-\frac{d}{2}+r}
\exp \Big( \frac{- {(M^{\pm}_{\pi})}^2}{4 \pi T^2} \rho \Big) \nonumber \\
& & \times \Big( \frac{1}{\sinh(\epsilon \rho/4 \pi)} - \frac{4 \pi}{\epsilon \rho} \Big) \, \Big[ S\Big(\frac{1}{\rho} \Big) -1 \Big] \, .
\end{eqnarray}
The crucial point is that -- in the chiral limit -- the mass $M^{\pm}_{\pi}$ of the charged pions does not tend to zero. Rather, according to
Eq.~(\ref{chargedPionMass}), a magnetic-field dependent mass term survives the chiral limit,
\begin{equation}
M^2_H = \frac{{\overline l}_6 - {\overline l}_5}{48 \pi^2} \, \frac{{|qH|}^2}{F^2}
= \frac{16 \pi^2}{3} \, ({\overline l}_6 - {\overline l}_5) t^4 F^2 {\epsilon}^2 \, ,
\end{equation}
with
\begin{equation}
t = \frac{T}{4 \pi F} \, .
\end{equation}
The pertinent expansion parameter in the weak magnetic field limit $|qH| \ll T^2 $ is
\begin{equation}
\epsilon = \frac{|qH|}{T^2} \, .
\end{equation}
To isolate divergences in the kinematical functions ${\tilde g}_r$ (where $r=0,1,2, \dots$) that arise in the limit $\epsilon \to 0$
($T$ held fixed while $H \to 0$), we decompose ${\tilde g_r}(M_H, T, H)$ into two pieces,
\begin{eqnarray}
\label{decompInt}
{\tilde g_r}(M_H, T, H) & = & \frac{\epsilon \, T^{d-2r}}{{(4 \pi)}^{r+1}} \, {\int}_{\!\!\! 0}^1 \mbox{d} \rho \, \rho^{-\frac{d}{2}+r}
e^{- \gamma \, \epsilon^2 \rho}  \, \Big( \frac{1}{\sinh(\epsilon \rho/4 \pi)} - \frac{4 \pi}{\epsilon \rho} \Big) \,
\Big[ S\Big(\frac{1}{\rho} \Big) -1 \Big] \nonumber \\
& & + \frac{\epsilon \, T^{d-2r}}{{(4 \pi)}^{r+1}} \, {\int}_{\!\!\! 1}^{\infty} \mbox{d} \rho \, \rho^{-\frac{d}{2}+r}
e^{- \gamma  \, \epsilon^2 \rho}  \, \Big( \frac{1}{\sinh(\epsilon \rho/4 \pi)} - \frac{4 \pi}{\epsilon \rho} \Big) \,
\Big[ S\Big(\frac{1}{\rho} \Big) -1 \Big] \nonumber \\
& = & I_a + I_b \, ,
\end{eqnarray}
where
\begin{equation}
\gamma = \frac{{\overline l}_6 - {\overline l}_5}{12 \pi} \, t^2  \, .
\end{equation}
The first integral $I_a$ exists for integer $r = 0,1,2, \dots$. Taylor expanding the integrand in the parameter $\epsilon$, we obtain a
series with ascending even powers of $\epsilon$ for $r = 0,1,2, \dots$,
\begin{equation}
{\alpha}_1 {\epsilon}^2 + {\alpha}_2 {\epsilon}^4 + {\alpha}_3 {\epsilon}^6 + {\cal O}({\epsilon}^8) \, .
\end{equation}
The explicit coefficients are irrelevant for our purposes because the respective terms do not contribute to the quark condensate at the
accuracy we are interested in (up to $\epsilon^2 \ln \epsilon $ in the finite-temperature quark condensate). In particular, no
$\epsilon$-divergences come from here.

We thus examine the second integral $I_b$ in Eq.~(\ref{decompInt}) that we process by using the Jacobi identity
\begin{equation}
S\Big( \frac{1}{z} \Big) = \sqrt{z} \, S(z) \, .
\end{equation}
We then obtain the three integrals
\begin{eqnarray}
\label{OneTwoThree}
I_b & = & \frac{\epsilon \, T^{d-2r}}{{(4 \pi)}^{r+1}} \, {\int}_{\!\!\! 1}^{\infty} \mbox{d} \rho \, \rho^{r-\frac{d}{2}+\frac{1}{2}}
e^{- \gamma \epsilon^2 \rho} \, \Big( \frac{1}{\sinh(\epsilon \rho/4 \pi)} - \frac{4 \pi}{\epsilon \rho} \Big) \, \Big[ S(\rho) -1
\Big] \nonumber \\
& & + \frac{\epsilon \, T^{d-2r}}{{(4 \pi)}^{r+1}} \, {\int}_{\!\!\! 1}^{\infty} \mbox{d} \rho \, \rho^{r-\frac{d}{2}+\frac{1}{2}}
e^{- \gamma \epsilon^2 \rho} \, \Big( \frac{1}{\sinh(\epsilon \rho/4 \pi)} - \frac{4 \pi}{\epsilon \rho} \Big)
\nonumber \\
& & - \frac{\epsilon \, T^{d-2r}}{{(4 \pi)}^{r+1}} \, {\int}_{\!\!\! 1}^{\infty} \mbox{d} \rho \, \rho^{r-\frac{d}{2}}
e^{- \gamma \epsilon^2 \rho}  \, \Big( \frac{1}{\sinh(\epsilon \rho/4 \pi)} - \frac{4 \pi}{\epsilon \rho} \Big) \nonumber \\
& = & I_{b1} + I_{b2} + I_{b3} \, .
\end{eqnarray}
The first one -- $I_{b1}$ -- exists for integer $r = 0,1,2, \dots$. Taylor expanding the integrand and then integrating term by term we get a
series of the form
\begin{equation}
{\beta}_1 {\epsilon}^2 + {\beta}_2 {\epsilon}^4 + {\beta}_3 {\epsilon}^6 + {\cal O}({\epsilon}^8) \, .
\end{equation}
Again, the coefficients are irrelevant at the accuracy we are interested in. To isolate potential $\epsilon$-divergences in $I_{b2}$, we
write the integration limits as
\begin{eqnarray}
\label{intLimits}
I_{b2} & = & \frac{\epsilon \, T^{d-2r}}{{(4 \pi)}^{r+1}} \, {\int}_{\!\!\! 0}^{\infty} \mbox{d} \rho \, \rho^{r-\frac{d}{2}+\frac{1}{2}}
e^{- \gamma \, \epsilon^2 \rho} \, \Big( \frac{1}{\sinh(\epsilon \rho/4 \pi)} - \frac{4 \pi}{\epsilon \rho} \Big) \nonumber \\
& & - \frac{\epsilon \, T^{d-2r}}{{(4 \pi)}^{r+1}} \, {\int}_{\!\!\! 0}^{1} \mbox{d} \rho \, \rho^{r-\frac{d}{2}+\frac{1}{2}}
e^{- \gamma \, \epsilon^2 \rho} \, \Big( \frac{1}{\sinh(\epsilon \rho/4 \pi)} - \frac{4 \pi}{\epsilon \rho} \Big) \, .
\end{eqnarray}
The first expression can be integrated analytically,
\begin{equation}
I^{[1]}_{b2} = T^{d-2r} \Gamma( r - \mbox{$\frac{3}{2}$} ) \Bigg[ -\frac{\gamma^{\frac{3}{2}-r}}{{(4\pi)}^r} \, \epsilon^{3-2r}
+ 2^{-r-\frac{5}{2}} \pi^{-\frac{3}{2}} (2r - 3) \zeta(r-\mbox{$\frac{1}{2}$}, \mbox{$ \frac{1}{2}$} + 2\pi \gamma \epsilon) \,
\epsilon^{\frac{3}{2}-r} \Bigg]\, ,
\end{equation}
where the generalized Riemann zeta function is defined as
\begin{equation}
\zeta(s,a) = \sum^{\infty}_{k=0} \, \frac{1}{{(k+a)}^s} \, .
\end{equation}
One notices that the integral $I^{[1]}_{b2}$ (for integer $r \ge 2$) leads to $\epsilon$-divergences in the functions ${\tilde g}_r$, namely
\begin{equation}
\label{leadingDivergences}
{\tilde g}_r \propto \frac{1}{\epsilon^{2r-3}} \, , \frac{1}{\epsilon^{r-\frac{3}{2}}} \, .
\end{equation}
As it turns out, these are indeed the leading divergences in the Bose functions ${\tilde g}_r$. With the second expression -- $I^{[2]}_{b2}$
-- in Eq.~(\ref{intLimits}) we proceed as before: Taylor expanding the integrand again gives rise to a series displaying even
$\epsilon$-powers whose respective coefficients are of no concern to us.

Finally, we analyze the remaining third integral $I_{b3}$ in Eq.~(\ref{OneTwoThree}). Regularizing it with $N \gg 1$,
\begin{equation}
I_{b3} = \lim_{N \to \infty} - \frac{\epsilon \, T^{d-2r}}{{(4 \pi)}^{r+1}} \, {\int}_{\!\!\! 1}^N \mbox{d} \rho \, \rho^{r-\frac{d}{2}}
e^{- \gamma \, \epsilon^2 \rho} \, \Big( \frac{1}{\sinh(\epsilon \rho/4 \pi)} - \frac{4 \pi}{\epsilon \rho} \Big) \, ,
\end{equation}
the substitution $z = \ln (\epsilon u)$ -- for the specific case $r$=2 (and $d$=4) -- leads to
\begin{equation}
I_{b3}(r=2) = \lim_{N \to \infty} \, -\frac{\epsilon^{-4 \pi \gamma \epsilon}}{16 \pi^2} \, {\int}_{\!\!\! u_0}^{u_N} \mbox{d} u
\, u^{-1-4 \pi \gamma \epsilon} \Big( \frac{1}{\sinh(\ln \epsilon u)} - \frac{1}{\ln \epsilon u} \Big) \, ,
\end{equation}
with
\begin{equation}
u_0 = \frac{e^{\frac{\epsilon}{4 \pi}}}{\epsilon} \, , \qquad u_N = \frac{e^{\frac{N}{4 \pi}}}{\epsilon} \, .
\end{equation}
The integral can be performed analytically,
\begin{equation}
\label{thirdAnalytic}
I_{b3}(r=2) = \frac{1}{16 \pi^2} \, \Bigg\{ {\cal B}\Big( e^{-\frac{N}{2\pi}}; \mbox{$ \frac{1}{2}$} + 2 \pi \gamma \epsilon, 0 \Big)
- {\cal B}\Big( e^{-\frac{\epsilon}{2\pi}}; \mbox{$\frac{1}{2}$} + 2 \pi \gamma \epsilon, 0 \Big)
- {\cal E}(- \gamma \epsilon^2) + {\cal E}(- \gamma N \epsilon) \Bigg\} \, ,
\end{equation}
where the incomplete beta function and the exponential integral function, respectively, are defined as 
\begin{eqnarray}
{\cal B}(z;a,b) & = & {\int}_{\!\!\! 0}^z \mbox{d} x \, x^{a-1} {(1-x)}^{b-1} \, , \nonumber \\
{\cal E}(z) & = & - {\int}_{\!\!\! -z}^{\infty} \mbox{d} x \, \frac{e^{-x}}{x} \, .
\end{eqnarray}
Expanding $I_{b3}$ in $\epsilon$, one notices that only the second and third expression in Eq.~(\ref{thirdAnalytic}) lead to
$\epsilon$-divergences. Concretely, we obtain a logarithmic divergence,
\begin{equation}
I_{b3}(r=2) = -\frac{1}{16 \pi^2} \ln \epsilon + {\cal O}(\epsilon^0) \, .
\end{equation}
Collecting results, the divergences in the function ${\tilde g}_2$ in the weak magnetic field limit are
\begin{equation}
{\tilde g}_2 = -\frac{1}{16 \pi^{\frac{3}{2}} \sqrt{\gamma}} \, \frac{1}{\epsilon}
-\frac{\sqrt{2}-4}{32 \pi} \, \zeta(\mbox{$\frac{3}{2}$}) \, \frac{1}{\sqrt{\epsilon}}
-\frac{1}{16 \pi^2} \ln \epsilon + {\cal O}(\epsilon^0) \, .
\end{equation}

The quark condensate in the chiral limit, according to Eqs.~(\ref{quarkCondensateTChiralLimit}) and (\ref{expansionMH}), features the
series
\begin{equation}
{\cal S}[\tilde g] = - {\hat c} \epsilon^2 {\tilde g}_2 + \frac{{\hat c}^2 \epsilon^4}{2!} \, {\tilde g}_3
- \frac{{\hat c}^3 \epsilon^6}{3!} \, {\tilde g}_4 + {\cal O}({\tilde g}_5) \, ,
\end{equation}
where
\begin{equation}
{\hat c} = 4 \pi T^2 \gamma \, , \qquad \gamma = \frac{{\overline l}_6 - {\overline l}_5}{12 \pi} \, t^2  \, .
\end{equation}
According to Eq.~(\ref{leadingDivergences}), the leading divergence in the functions ${\tilde g}_r$ is proportional to $\epsilon^{3-2r}$.
Therefore each term in the above series gives rise to a contribution linear in $\epsilon$. All these terms have to be taken into account at
the order we are operating. The series can be resummed with the result
\begin{equation}
\label{resumTildeg}
{\cal S}[\tilde g] = \frac{\sqrt{2}-1}{2 \sqrt{\pi}} \, \sqrt{\gamma} \, \epsilon \, T^2 \, .
\end{equation}
The quark condensate in the chiral limit -- see Eqs.~(\ref{quarkCondensateTChiralLimit}) and (\ref{expansionMH}) -- furthermore involves
the other type of Bose functions $g_r(M,T,0)$. The structure of the expansion in the mass parameter $M$ for the specific function
$g_0(M,T,0)$ has been analyzed in Refs.~\citep{GL89,Hof10} with the outcome
\begin{eqnarray}
g_0(M,T,0) & = & T^4 \, \Bigg[ \frac{\pi^2}{45} \, - \, \frac{1}{12}
\frac{M^2}{T^2} \, + \, \frac{1}{6 \pi} \frac{M^3}{T^3}
\, + \,  \frac{ (2\gamma_E -\mbox{$\frac{3}{2}$} )}{32{\pi}^2} \frac{M^4}{T^4} \, 
+ \frac{1}{32{\pi}^2} \, \frac{M^4}{T^4} \, \ln \frac{M^2}{16 \pi^2 T^2}
\nonumber \\
& & + 2 \pi^{3/2} \, \sum_{n=3}^{\infty} \frac{(-1)^n}{n!} \,
\Big(  \frac{M}{2 \pi T} \Big)^{2n} \, \Gamma(n-\mbox{$\frac{3}{2}$}) \,
\zeta(2n-3) \Bigg] \quad (T \gg M) \, .
\end{eqnarray}
With the recursion relation
\begin{equation}
g_{r+1} = - \frac{\mbox{d} g_r}{\mbox{d} M^2} \, ,
\end{equation}
the series for any other $g_r$ with $r =1,2,3, \dots$ can be derived.

In our case of interest, the relevant mass in these functions is $M_H$, 
\begin{equation}
M^2_H = \frac{{\overline l}_6 - {\overline l}_5}{48 \pi^2} \, \frac{{|qH|}^2}{F^2}
= \frac{16 \pi^2}{3} \, ({\overline l}_6 - {\overline l}_5) t^4 F^2 {\epsilon}^2 \, ,
\end{equation}
i.e., the mass of the charged pion that survives the chiral limit. We then find that the leading $\epsilon$-divergences in these functions
are
\begin{equation}
g_r = \frac{(2r-5)!! \, \gamma^{3/2-r}}{2^{3r-2} \pi^{r-1/2}} \, \epsilon^{3-2r} \, , \qquad r=2,3,\dots \, .
\end{equation}
The series of kinematical functions $g_r$, as it occurs in the quark condensate,
\begin{equation}
{\cal S}[g] = - {\hat c} \epsilon^2 g_2 + \frac{{\hat c}^2 \epsilon^4}{2!} \, g_3 - \frac{{\hat c}^3 \epsilon^6}{3!} \, g_4 + {\cal O}(g_5)
\, ,
\end{equation}
hence yields an infinite number of terms that are all linear in $\epsilon$. Resumming, we obtain
\begin{equation}
{\cal S}[g] = \frac{1 - \sqrt{2}}{2 \sqrt{\pi}} \, \sqrt{\gamma} \, \epsilon \, T^2 \, .
\end{equation}
This just cancels the contribution from ${\cal S}[\tilde g]$, Eq.~(\ref{resumTildeg}), such that there are no terms linear in $\epsilon$ in
the quark condensate coming from here. The logarithmic contributions, however, that are present both in $g_2$ and ${\tilde g}_2$ do not
cancel: in the sum we have
\begin{equation}
g_2 + {\tilde g}_2 = \frac{1}{16 \pi^2} \ln \epsilon + {\cal O}(\epsilon^{-1/2}) \, ,
\end{equation}
giving rise to a contribution $\epsilon^2 \ln \epsilon$ in the quark condensate.

Finally, the $\epsilon$-expansion for the functions $g_1(0,T,0)$ and ${\tilde g}_1(0,T,H)$ that also appear in the quark condensate,
Eqs.~(\ref{quarkCondensateTChiralLimit}) and (\ref{expansionMH}), has been provided in Refs.~\citep{GL89,Hof19}. For completeness we quote
the result,
\begin{eqnarray}
g_1(0, T, 0) & = & \frac{1}{12} \, T^2 \, , \\
{\tilde g_1}(0, T, H) & = & - \Bigg\{ \frac{|I_{\frac{1}{2}}|}{8 \pi^{3/2}} \sqrt{\epsilon}
- \frac{\ln 2}{16 \pi^2} \, \epsilon
+\frac{\zeta(3)}{384 \pi^4} \, \epsilon^2
- \frac{7 \zeta(7)}{98 304 \pi^8} \, \epsilon^4
+ {\cal O}(\epsilon^6) \Bigg\} \, T^2 \, , \nonumber
\end{eqnarray}
with
\begin{equation}
\label{I12}
I_{\frac{1}{2}} = {\int}_{\!\!\! 0}^{\infty} \, \mbox{d} \rho \rho^{-1/2} \Big( \frac{1}{\sinh(\rho)} - \frac{1}{\rho} \Big) \approx -1.516256 \, .
\end{equation}

\end{appendix}

\end{document}